\documentclass[preprint,prb,superscriptaddress,byrevtex,showpacs, 
showkeys]{revtex4}
\usepackage{amsmath}
\usepackage{amsfonts}

\input{tcilatex}
\begin{document}

\preprint{}
\title[Short title for running header]{Magnetic behavior of Ba$_{3}$Cu$_{3}$%
Sc$_{4}$O$_{12}$}
\author{B. Koteswararao}
\affiliation{Department of Physics, Indian Institute of Technology Bombay, Mumbai 400076,
India}
\author{A. V. Mahajan}
\affiliation{Department of Physics, Indian Institute of Technology Bombay, Mumbai 400076,
India}
\author{F. Bert}
\affiliation{Laboratoire de Physique des Solides, Universit\'{e} Paris-Sud, UMR8502,
CNRS, 91405 Orsay Cedex, France.}
\author{P. Mendels}
\affiliation{Laboratoire de Physique des Solides, Universit\'{e} Paris-Sud, UMR8502,
CNRS, 91405 Orsay Cedex, France.}
\author{ J. Chakraborty}
\affiliation{Department of Solid State Physics, Indian Association for the Cultivation of
Science, Jadavpur, Kolkata 700 032, India}
\author{V. Singh}
\affiliation{Department of Physics, Indian Institute of Technology Bombay, Mumbai 400076,
India}
\affiliation{Department of Solid State Physics, Indian Association for the Cultivation of
Science, Jadavpur, Kolkata 700 032, India}
\author{I. Dasgupta}
\affiliation{Department of Physics, Indian Institute of Technology Bombay, Mumbai 400076,
India}
\affiliation{Department of Solid State Physics, Indian Association for the Cultivation of
Science, Jadavpur, Kolkata 700 032, India}
\author{S. Rayaprol}
\affiliation{UGC-DAE Consortium for Scientific Research, Mumbai Centre, R-5 Shed, Bhabha
Atomic Research Centre, Mumbai 400 085, India}
\author{V. Siruguri}
\affiliation{UGC-DAE Consortium for Scientific Research, Mumbai Centre, R-5 Shed, Bhabha
Atomic Research Centre, Mumbai 400 085, India}
\author{A. Hoser}
\affiliation{Helmholtz-Zentrum Berlin f\"{u}r Materialien und Energie GmbH,\\
Hahn-Meitner-Platz 1\\
D-14109 Berlin, Germany}
\author{S. D. Kaushik}
\affiliation{UGC-DAE Consortium for Scientific Research, Mumbai Centre, R-5 Shed, Bhabha
Atomic Research Centre, Mumbai 400 085, India}

\begin{abstract}
The chain-like system Ba$_{3}$Cu$_{3}$Sc$_{4}$O$_{12}$ has potentially
interesting magnetic properties due to the presence of Cu$^{2+}$ and a
structure-suggested low-dimensionality. \ We present magnetization $M$
versus magnetic field $H$ and temperature $T$, $T$- and $H$-dependent
heat-capacity $C_{p}$, $^{45}$Sc nuclear magnetic resonance (NMR), muon spin
rotation ($\mu $SR), neutron diffraction measurements and electronic
structure calculations for Ba$_{3}$Cu$_{3}$Sc$_{4}$O$_{12}$. \ \ The onset
of magnetic long-range antiferromagnetic order at $T_{\text{N}}$ $\sim 16$ K
is consistently evidenced from the whole gamut of our data. \ A significant
sensitivity of $T_{\text{N}}$ to the applied magnetic field $H$ ($T_{\text{N}%
}\sim 0$ K for $H=70$ kOe) is also reported. \ Coupled with a ferromagnetic
Curie-Weiss temperature ($\theta _{\text{CW}}\sim 65$ K) in the
susceptibility (from a 100 K-300 K fit), it is indicative of competing
ferromagnetic and antiferromagnetic interactions. \ These indications are
corroborated by our density functional theory based electronic structure
calculations where we find the presence of significant ferromagnetic
couplings between some copper ions whereas AF couplings were present between
some others. \ Our experimental data, backed by our theoretical
calculations, rule out one-dimensional magnetic behaviour suggested by the
structure and the observed long-range order is due to the presence of
non-negligible magnetic interactions between adjacent as well as
next-nearest chains.
\end{abstract}

\volumeyear{year}
\volumenumber{number}
\issuenumber{number}
\eid{identifier}
\date[Date text]{date}
\received[Received text]{date}
\revised[Revised text]{date}
\accepted[Accepted text]{date}
\published[Published text]{date}
\startpage{1}
\endpage{2}
\maketitle

\section{\textbf{Introduction}}

One dimensional magnetic systems are recognized as an ideal playground for
the search of novel states of matter. Indeed, by varying the interactions
(near neighbour, next nearest neighbour) various states can be stabilized
and have been the focus of experimental studies in the recent past, where
high field magnetization experiments have played a central role. Their
experimental exploration has in turn induced some theoretical developments
to understand the observed properties. Various models for one-dimensional
magnetic systems with different spin Hamiltonians have been theoretically
investigated. \ It is only in the last 10-15 years that experimentalists are
coming up with actual realizations of different model systems. \ Clearly,
real systems are quite often much more complex compared to simple models. \
In such cases, further efforts by theorists based on information pertaining
to a real system, often lead to an improved understanding of its properties.
Among the various chain-like $S=%
%TCIMACRO{\U{bd}}%
%BeginExpansion
{\frac12}%
%EndExpansion
$, antiferromagnetic (AF) systems investigated, a compound with a
"diamond-chain" geometry has been under focus recently. The system consists
of isolated chains of corner-shared diamonds (squares). With the presence of
antiferromagnetic couplings and possibility of frustrated nearest-neighbour
interactions (due to the two edge-shared triangles within a diamond), exotic
physics might result. In practice, Cu$_{3}$(CO$_{3}$)$_{2}$(OH)$_{2}$ or
azurite is the only compound investigated with the diamond chain structure.
Measurements of the susceptibility, heat-capacity, and high-field
magnetization in this compound have revealed it to be a prototypical
substance for the frustrating diamond-chain model.\cite{H. Kikuchi 2005}
High-field magnetization data evidence the presence of a $1/3$ magnetization
plateau. If one considers the chain to \ comprise of alternating dimers and
monomers, this plateau is thought to arise due to complete polarization of
the monomers by the applied magnetic field. As for the correct model which
could describe the experimental data, the initial work suggested all the
interactions to be antiferromagnetic. However, inelastic neutron scattering
experiments \cite{K.C.Rule 2008} actually suggested that one of the
interactions between the spins comprising the dimer and the monomer was
ferromagnetic. \ In contrast, later numerical studies \cite{H.Jeschke} found
that all the experimental data could be explained by considering a model
where all the interactions were antiferromagnetic (and therefore strongly
frustrated) and there also existed an interaction between the monomers.
Motivated by this work, we thought it would be interesting to search for
other systems with a similar geometry, the studies on which might help to
elucidate the physics behind the unusual features of azurite. One such
possible system could be Ba$_{3}$Cu$_{3}$Sc$_{4}$O$_{12}$ (BCSO)\textbf{.}

At first sight, the structure of BCSO is quite appealing. \ Consider the CuO$%
_{4}$ plaquette as a unit. \ Now imagine a square made up of four such units
with Cu at the vertices. \ If we arrange these squares in a corner-sharing
configuration, we will get the classic diamond-chain. \ In BCSO, the only
difference with respect to the classic diamond-chain is that the adjacent
squares have their planes orthogonal to each other (rotated about the
diagonal). \ \ Given that there has been hardly any experimental/theoretical
work pertaining to BCSO except for structural studies \cite{Gregory JMC 2001}
and a nuclear magnetic resonance (NMR) investigation, \cite{AAGipplus1989}
BCSO therefore appears to be a fit case for further investigation. \ There
are two important issues that need to be addressed. \ (i) Are the magnetic
linkages as simple as suggested by the structure? The magnetism would
therefore be that of a low-dimensional compound. (ii) What are the dominant
magnetic interactions and what are their relative magnitudes and signs?

With the above considerations in mind we have undertaken a detailed
investigation of BCSO. \ Bulk techniques such as susceptibility, heat
capacity and neutron diffraction have been used to establish the gross
nature of magnetism in the material. \ Local probe $^{45}$Sc NMR has helped
us to determine local magnetic and structural properties via a measurement
of the $^{45}$Sc NMR shift and the line shape. \ The main outcome of the
above measurements is that BCSO does not show any signs of
low-dimensionality in the temperature range investigated 1.8 K-300 K. \ On
the contrary, clear evidence of AF order at $T_{\text{N}}$ = 16 K from the
bulk ($\sim $100\% as shown from zero-field $\mu $SR) of the sample is seen
while susceptibility data at high-$T$ (100 K-300 K) are indicative of
ferromagnetic interactions. One of the goals of the present paper\textbf{\ }%
is to reconcile this apparent contrast between the expectations based on
atomic arrangements and the observed LRO and to provide a basis for
understanding\textbf{\ }ferromagnetic interactions at high-$T$ but AF order
at low-$T$. \ In order to address these issues, we have carried out density
functional theory based electronic structure calculations. \ We have as well
calculated various exchange interactions in the Generalised Gradient
Approximation GGA\ +\ U\ framework (here, U is the on-site Coulomb repulsion
energy). \ The results of these calculations enable us to resolve the
apparent surprises of the experimental data. \ The main conclusion is that
the strongest magnetic interaction in BCSO is between a Cu and its nearest
neighbour and is ferromagnetic. \ The diagonal interactions (in the diamond)
as also the next-nearest neighbour interactions along the chain are AF and
substantial (about half the magnitude of the nearest neighbour couplings). \
The observed LRO appears to be due to comparably strong couplings between
neighboring chains. Due to the frustration generated by a mixture of
comparable F and AF couplings, the system is susceptible to loss of LRO in
moderate applied magnetic fields.

The paper is organized as follows. \ Following the section on experimental
details, we present the results of our various measurements corresponding to
the paramagnetic state of BCSO. \ While the x-ray diffraction and neutron
diffraction data validate the structure of \ BCSO, bulk susceptibility and
heat capacity help characterize the magnetic interactions and the lattice
properties. We next discuss results which establish the onset of LRO in BCSO
at $T$ $\sim $16 K. \ This is followed by presentation of the results for
the $T$-region below the ordering temperature including the effect of an
applied magnetic field on the ordering temperature. \ This is followed by a
discussion section which uses the results of our electronic structure
calculations which are instrumental in obtaining a coherent understanding of
the experimental data. \ The important conclusions are given in the last
section.

\section{\textbf{Experimental details}}

The polycrystalline samples of Ba$_{3}$Cu$_{3}$Sc$_{4}$O$_{12}$ were
synthesized by a solid state reaction route with stoichiometric amounts of
BaCO$_{3}$ $(99.99\%)$, CuO $(99.7\%)$, and Sc$_{2}$O$_{3}$ $(99.99\%)$.
These amounts were ground, pelletized, and fired in a alumina crucible at $%
850^{\circ }$C to $1050^{\circ }$C in air for $4$ days with regrinding every 
$24$ hours. Samples were quench-cooled to room temperature in air. Powder
x-ray diffraction data were collected using a PANalytical X'pert PRO powder\
diffractometer with Cu K$\alpha $ radiation. The Rietveld refinements were
carried out using Fullprof software, \cite{fullprof} and confirmed the
single-phase nature of the sample. Magnetization $(M)$\ measurements were
done in the temperature $(T)$ range from 2 K to 300 K and in the\ magnetic
field $(H)$ range 0-90 kOe using a SQUID magnetometer (MPMS) and a vibrating
sample magnetometer\ VSM (PPMS), both from Quantum Design. Heat-capacity
measurements as a function of $T$ and $H$ were also performed using the PPMS
by the thermal\ relaxation method. Further, we have carried out $^{45}$Sc
NMR spectra measurements in the $T-$range from $1.2$ K to $80$ K using the
spectrometer attached with a magnet having a sweepable field. \ This was
done at various fixed frequencies. The spectra were measured from\ $80$ K to 
$300$ K using the spectrometer attached with a fixed field of about $70$
kOe. The preparation and bulk measurements were done at IIT Bombay while the
NMR work was done at the Laboratoire de Physique des Solides (LPS), Orsay. $%
\ $The $\mu SR$ measurements were carried out to probe the ordering and were
performed at Paul Scherrer Institut (PSI), Villigen (Switzerland). Neutron
diffraction (ND) measurements were carried out on powder samples using the
focusing crystal diffractometer (FCD) set up by UGC-DAE Consortium for
Scientific Research Mumbai Centre at the Dhruva reactor, Mumbai (India) at a
wavelength of 1.48 \AA\ (\cite{siruguri}) and the diffractometer E6
installed at the BERII reactor (Helmholtz-Zentrum Berlin) at a wavelength of
2.45 \AA . For both experiments, powder sample was filled in a vanadium can.
\ ND experiments were carried out in the temperature range 2 K to 300 K and
also under external magnetic fields up to 80 kOe. For in-field experiments,
a mixture of deuterated methanol and ethanol was used as a binder to avoid
preferred orientation of samples due to the magnetic field. \ ND patterns
were analyzed by the Rietveld method using the program FullProf for the
refinement of chemical and magnetic structures. The results of our
measurements and calculations will be described in the following sections.

\section{\textbf{Experimental results and analysis}}

Ba$_{3}$Cu$_{3}$Sc$_{4}$O$_{12}$ crystallizes with a tetragonal structure in
the space group I4/mcm (No. 140). The lattice parameters obtained by us are $%
a=11.9021$ \AA , $c=8.3962$ \AA , $z=4$, and $c/a=0.71,$ which are in good
agreement with previously reported values.\cite{Gregory JMC 2001}

The structure of Ba$_{3}$Cu$_{3}$Sc$_{4}$O$_{12}$ was first reported by
Duncan H. Gregory.\cite{Gregory JMC 2001} The crystal structure is shown in
Fig. $1$. In this figure, the chain formed by Cu and O atoms (enclosed by
blue dashed lines with an arrow mark) is shown separately. The chains
consist of diamond (or square) like units, which are built from \ two types
of CuO$_{4}$\ units (Cu(1)O$_{4}$ and Cu(2)O$_{4}$), as shown in Fig. $1$.
The\ Cu atoms in the square are coupled to their nearest neighbors via
oxygen with a Cu-O-Cu bond angle of $\sim $ 90$^{o}$. Successive\ Cu-squares
(formed by 4 Cu atoms) are corner shared with their planes perpendicular to
each other. The chains are separated by Ba atoms. Looking at these figures,
one can get the impression that there exist diamond-like chains in the $c-$%
direction.

\subsection{Probing the paramagnetic state and the ordering using bulk probes%
}

\subsubsection{Magnetisation}

In Fig. 2(a), $\chi (T)=M/H$ \ increases with decreasing $T$ in a
Curie-Weiss manner. Although the structure is suggestive of a
one-dimensional spin system, no broad maximum was seen in $\chi (T)$. The
inverse of $\chi (T)$ versus $T$ is also\ plotted in the right $y-$axis of
the Fig. 2(a). The high temperature (200 K-300 K) $\chi (T)$ is well
described by the Curie-Weiss law $\chi (T)=\chi _{0}+C/(T-\theta _{CW})$
with $\chi _{0}=-(7\pm 1)\times 10^{-5}$ cm$^{3}$/mol Cu, $C=0.413$ cm$^{3}$
K/mol Cu (fixed to this value corresponding to $S=%
%TCIMACRO{\U{bd}}%
%BeginExpansion
{\frac12}%
%EndExpansion
$ and $g=2.1$), and $\theta _{CW}=65\pm 2$ K$.$ The positive value of
Curie-Weiss temperature $\theta _{CW}=65\pm 2$ K suggests that the major
exchange couplings are ferromagnetic at high$-T$.\ Here $\chi _{0}$ is the
sum of diamagnetic susceptibility $\chi _{core}$ and Van Vleck
susceptibility $\chi _{vv}$. The core diamagnetic susceptibility is
calculated to be about $-9.63\times 10^{-5}$ cm$^{3}$/mol Cu.\cite{P.W.
Selwood} This yields $\chi _{vv}\sim (2.6\pm 1.0)\times 10^{-5}$ cm$^{3}$%
/mol Cu which is comparable with other cuprates. \ The sharp maximum in the
susceptibility at about $16$ K\ indicates AF LRO in this system. The ratio $%
\theta _{CW}/T_{N}\sim $ $4$, admits the possibility of a moderate level of
frustration in our system. The presence of ferromagnetic interactions at
high temperature and antiferromagnetic ordering at low temperature in this
sample is interesting to motivate further measurements.

To understand the nature of ordering in the sample, we have measured $\chi
(T)$ in various $H$ of $100$ Oe, $20$ kOe, $30$ kOe, $40$ kOe, $60$ kOe, and 
$80$ kOe in the $T-$range $2$ to $50$ K as shown in Fig. $2(b)$. There is no
difference found in the data of zero field cooling (ZFC) and field cooling
(FC) $\chi (T)$ measured at $100$ Oe. This indicates that the system doesn't
have disordered or spin-glass like behavior. The $T_{N}$ decreases with $H$,
which is also a clear indication that the ordering is of the
antiferromagnetic kind. Surprisingly, $T_{N}$\ vanishes even at a relatively
small field at about $60-80$ kOe. In a field of $60$ kOe, the sharp peak in $%
\chi (T)$ disappears and $\chi (T)$ is nearly $T-$independent. The $\chi (T)$
at 80 kOe doesn't have any anomaly. Magnetization $M$ data as a function of $%
H$ up to $90$ kOe at different $T$'s are shown in Fig. $3$. The obtained $%
T_{N}$'s from $\chi (T)$ are plotted in Fig. $5$. In the case of $M(H)$ at $%
2 $ K, $M$ increases with\ field and saturates to about $1$ $\mu _{B}$ above 
$80$ kOe, as shown in the inset of\ Fig. $3$. The data are also compared
with magnetization as a function of applied field at $T=2$ K generated using
the Brillouin function for isolated $S=%
%TCIMACRO{\U{bd}}%
%BeginExpansion
{\frac12}%
%EndExpansion
$ moments with $g=2.1$ (typical for cuprates)$.$ The data at low fields have
values which are smaller than the free spin behavior due to
antiferromagnetic coupling between the spins at $T=2$ K. At high fields
above $80$ kOe, the data follow the $S=%
%TCIMACRO{\U{bd}}%
%BeginExpansion
{\frac12}%
%EndExpansion
$ $(g=2.1)$ Brillouin function. A small change in the slope of the
magnetization is observed at $2$ K at $H=23$ kOe. The change can be clearly
seen in the derivative of magnetization with field.\ This probably arises
from a spin-flop transition.

\subsubsection{Heat capacity}

In order to understand the field sensitive ordering in this system, we
performed heat-capacity measurement in different fields upto $90$ kOe using
the heat-capacity set-up (attached to the PPMS) by the thermal\ relaxation
method. The Fig. 4(a) shows $C_{p}$ versus $T.$ The $C_{p}$ has two
contributions, lattice and magnetic. In order to separate the lattice
contribution, it is normally useful to make measurements on an isostructural
non-magnetic analog of the given system (possibly by replacing the Cu atoms
by Zn and Mg atoms). However, neither is there is any report of a
non-magnetic analog of BCSO nor did we succeed in preparing such a compound.
We then went ahead and analyzed the lattice part of the $C_{p\text{ }}$data
in the $100-200$ K region using the Debye model.\cite{kittel} The data could
not be fit with a\ single Debye temperature, but we were able to fit the
data to the following formula in the $T-$range $100$ K to $200$ K, which
contains a linear combination of two Debye integrals.

\begin{equation}
C_{p}(T)=9rNk_{B}\sum\limits_{i=1,2}C_{i}\left( \frac{T}{\theta _{D}^{i}}%
\right) ^{3}\int\limits_{0}^{x_{D}^{i}}\frac{x^{4}e^{x}}{(e^{x}-1)^{2}}dx
\end{equation}

Here $r$ is the number of atoms per formula unit, $\theta _{D}^{i}$ is a
Debye temperature and $x_{D}^{i}=\theta _{D}^{i}/T$. The\ obtained
parameters are $C_{1}=0.61\pm 0.05,$ $\theta _{D}^{1}=(736\pm 10)$ K$,$ $%
C_{2}=0.33\pm 0.05,$ and $\theta _{D}^{2}=230\pm 30$ K. Although the fit is
good, the magnetic heat-capacity, obtained from $C_{p}$ by subtracting the
lattice contribution, has a broad maximum at about $30$ K which seems
unphysical. The entropy change calculated from magnetic specific heat is
about $2.6$ J/mol K Cu, which is smaller than that expected ($\sim R\ln $2)
for $S=%
%TCIMACRO{\U{bd}}%
%BeginExpansion
{\frac12}%
%EndExpansion
$ systems. Extracting the lattice contribution by fitting to a Debye model
in a lower range (say 40 - 100 K) might help to remove the 30 K broad
anomaly. However, the entropy change in that case would be even smaller.
This suggests the existence of a magnetic contribution to heat-capacity at
temperatures much higher than $T_{N}$ (the deviation from CW law in the
magnetic susceptibility begins above 100 K). \ This is in tune with our
electronic structure calculations (presented in a latter section) which
suggest the existence of long-range ferromagnetic couplings.

A sharp anomaly is observed at$\ 16$ K (as in $\chi (T)$)\ which is a clear
indication of ordering. In Fig. $4(b)$, $C_{p}/T$ is plotted as a function
of $T$ for various magnetic fields. The peak moves to lower temperatures as
the applied field is increased until it is smeared out and more-or-less
disappears when the applied field is $80$ kOe. The $T_{N}$'s (peak
positions) obtained from $C_{p}/T$ vs $T$ data are in agreement with those
obtained from $\chi (T)$ data (see Fig. $5$). In addition to the $16$ K
anomaly (in zero field) there is also a broad maximum around $3$ K in the $%
C_{p}/T$ vs $T$ data. While this, in principle, could be due to some change
in the spin-order, we did not find any evidence of this in our $\mu $SR data
down to $1.5$ K. Another possibility is that it is a Schottky anomaly from $%
S=%
%TCIMACRO{\U{bd}}%
%BeginExpansion
{\frac12}%
%EndExpansion
$ defects. However, no significant field-dependence is seen for the low$-T$
anomaly. \ But this could be simply because it gets masked by the large
contribution from the heat capacity from the AF part. For the same reason,
any Curie-like upturn in the low$-T$ susceptibility is overwhelmed by the
large susceptibility of the AF part.

\subsubsection{Neutron diffraction}

Fig. $6(a)$ shows the refinement of the ND pattern at $300$ K. Cell
parameters obtained by ND are in fair agreement with the XRD results. The
results of the refinement are given in Table $1$ along with the various
R-factors. \ Fig. $6(b)$ shows the ND patterns as a function of temperature
up to $30$ K. A weak low angle magnetic reflection at $11.8$ degree is
clearly seen at $2$ K, which gradually broadens and disappears by about $20$
K. The propagation vector for the refinement of the magnetic cell was
identified by using the $k$-search program within Fullprof suite. The result
of $k$-search gave a commensurate propagation vector $k=(010)$. Refinement
of ND pattern at 2 K was done by including a magnetic phase and the
propagation vector, $k=(010)$. For determining the magnetic structure, a
triclinic space group, I --1 was used to generate full set of reflections
with proper multiplicity. The irreducible representations of the little
groups were determined using BasIreps program within the Fullprof suite. The
upper tick marks belong to the nuclear phase while the lower ones are for
the magnetic phase. Inset $(i)$ of Fig. $6(b)$ shows the integrated
intensity of the $(0$ $0$ $0)$ magnetic reflection. There is no significant
change observed\ in the structural parameters obtained at $300$ K and $2$
K.\ The moment values on Cu1 and Cu2 obtained at $T=2$ K\ are about -0.65$\ $
$\mu _{B}$ and 1.15 $\mu _{B}$, respectively. Inset $(ii)$ of Fig. $6(b)$
shows the variation of moments of these two atoms with temperature. The ND
patterns taken at $2$ K under external magnetic fields up to $80$ kOe are
shown in Fig. $7$. The low angle magnetic reflection $(0$ $0$ $0)$ gets
completely suppressed under fields of $60$ kOe and above, indicating a
suppression of magnetic ordering. Considering that only one magnetic Bragg
peak was clearly observed and the associated uncertainties in determining
the integrated peak intensities (in the presence of a background), we feel
that the obtained magnetic moment values are in tune with those from our
bulk magnetisation data in high fields.

\bigskip

The above results then clearly suggest that the BCSO system must have
significant three-dimensional magnetic interactions which lead to the
observed LRO. \ Further, our observations indicate the presence of competing
magnetic \ interactions/frustrations. \ Both these suggestions are later
validated by our electronic structure calculations.

\subsection{\protect\bigskip Local probe measurements of the paramagnetic
state and the ordered state}

While the bulk probe data provide important input concerning the physical
properties of the system, local probes such as $\mu $SR and NMR enable us to
obtain a deeper look at the system often giving fine details which are not
accessible otherwise. \ 

\subsubsection{Muon spin rotation ($\protect\mu $SR)}

Strong evidence of a conventional magnetic ordering is provided by $\mu $SR
data taken in zero applied field. Indeed, as reported in Fig.~8, clear
oscillations of the time evolution of the muon polarization (the
"asymmetry"of the muon decay~\cite{musr-gen}) are observed at 1.5~K. The
signal combines several frequencies, as seen from the Fourier transform
spectra of Fig.~9. This is a clear-cut signature of the existence of well
defined internal fields on several inequivalent muon stopping sites. These
local fields cause a coherent precession of the muon spin components
perpendicular to the local field direction, averaging to a 2/3 value of the
total spin polarization for a powder sample. These well defined oscillations
are therefore characteristic of a magnetic long-range ordered state at low
temperature. \ Also, the magnetic order arises from the bulk (nearly 100 \%)
of the sample. The precession frequency $\nu _{i}$ is given by $\nu
_{i}=\gamma _{\mu }/2\pi \times B_{i}$ where $\gamma _{\mu }/(2\pi )=135.5$
MHz/T and $B_{i}$ is the average magnitude of the local field at the \textit{%
i}th muon site. In addition, the long-time asymmetry in our background-free
experiment tends to the remaining non-oscillating 1/3 part of the total
asymmetry which, in a static picture, accounts for the powder averaged
fraction of the muon spin components that lie parallel to the local magnetic
field. This 1/3 value is therefore typical of the absence of strong
dynamical effects in this system. The existence of multiple sites is in tune
with the structure since there are 3 inequivalent oxygen sites and, usually,
in oxides, muons bind to an oxygen at a distance $\sim 1$ \AA . \ Note that
there can be more magnetically inequivalent muon trapping sites than the
number of inequivalent oxygen ions.

Typical spectra in a narrow $T$-range around the magnetic transition are
displayed in the bottom panel of Fig.~8. A slowly relaxing Gaussian
asymmetry originating solely from quasi-static nuclear dipolar fields,
dominates the paramagnetic regime above the transition temperature (see the
16.58~K data). It is found to sharply replace the oscillating asymmetry in
the selected narrow $T$-range. This underlines the sharpness of the
transition which can be estimated to be at $T_{N}=16.2(1)$~K from this plot,
in fair agreement with that extracted from the zero field susceptibility and
heat-capacity data. The $T$-variations of the frequency of the oscillations
can be followed up to the magnetic transition (Fig.~9) and scale with each
other. This shows that once the order is established at $T_{N}$ no further
variation (even minute) of the magnetic structure occurs down to the lowest
probed temperature.

After normalizing the frequencies (these track the order parameter) to their
value at $T=1.5$~K, we find that their temperature variation (in the
temperature range above 10 K) obeys the following equation 
\begin{equation*}
\nu (T)=\nu (T=0)[1-(T/T_{N})]^{\beta }
\end{equation*}%
, with $\beta =0.26(2)$, and $T_{N}=16.2(1)$. This exponent is less than
that expected for three-dimensional models $\beta =0.367$ for the 3D
Heisenberg model, but larger than the one expected for 2D XY models $\beta
=0.23$ ~\cite{Bramwell}.

\subsubsection{Nuclear magnetic resonance (NMR)}

NMR measurements were carried out in the $T-$range from $1.2$ K to $300$ K
in the\ phase coherent spectrometers at LPS (Orsay) attached to fixed-field
and sweepable-field magnets. In these experiments, $^{45}$Sc nuclei (nuclear
spin $I=7/2$ and gyromagnetic ratio $\gamma /2\pi =10343.15$ kHz/T) were
probed using the pulsed NMR technique. In the structure, $^{45}$Sc nuclei
are in an\ octahedral ScO$_{6}$ environment with bond lengths as Sc-O1$\sim
2.04$ \AA , Sc-O2$\sim 2.11$ \AA , and Sc-O3$\sim 2.18$ \AA , respectively.
Since there is a slight\ deviation from a perfect octahedral symmetry, it
may result in quadrupole broadened NMR spectra.

\paragraph{Spectra at high-T (above 80 K):}

Fourier transform spectra were obtained in a fixed field of $\sim 70$ kOe in
the $T-$range from $80$ K to $300$ K (see Fig. $10$). Normalized
intensities\ are plotted as a function of normalized frequency $(\frac{\nu
_{o}-\nu _{ref}}{\nu _{0}}),$ where $\nu _{ref}=\frac{\gamma }{2\pi }%
H=72350.3$ kHz. In the case of single crystals, when quadrupole effects are
present ($\nu _{Q}\neq 0$) and there is uniaxial symmetry ($\eta =0$), the
peak position gives $K_{z}$. Since $^{45}$Sc nuclei in Ba$_{3}$Cu$_{3}$Sc$%
_{4}$O$_{12}$ are not in an environment of perfect octahedral symmetry, a
quadrupole-broadened powder pattern is expected with characteristic steps
and singularities. Note that the satellite intensities (Fig. 10(a) inset)
are smaller than those expected in a quadrupole powder pattern. This means
that many sites have a nearly octahedral symmetry and a small $\nu _{Q}$ and
the satellites come from a fraction of sites which deviate from the
octahedral symmetry due to distortions. \ The observed spectrum at 280 K can
be generated by a combination of a quadrupole powder pattern (spectrum A)
with $\eta \sim 0.25$ and $\nu _{Q}\sim 400$ kHz and another one with a
coincident central line (spectrum B) but $\nu _{Q}\sim 0$ (see inset of Fig.
10 (a)).

The peak position changes with $T$ and tracks the susceptibility ($N_{A}\mu
_{B}K=A_{hf}\chi _{spin}$). The shift is calculated from the position of the
maximum as from the formula $K_{z}=\frac{\nu _{o}^{\max }-\nu _{ref}}{\nu
_{ref}}.$ The obtained shifts as a function of spin susceptibility $\chi
_{spin}$ $(=\chi (T)-\chi _{o})$ are plotted in the inset of Fig. $10$ (b)
The slope from a linear fit gives the total hyperfine coupling constant ($%
A_{hf}$) between Sc and surrounding Cu atoms to be about $-4.5$ kOe. The
chemical shift obtained from the intercept of the fit is nearly zero. In
Ref. [\cite{AAGipplus1989}] the data were analyzed by plotting $K$ as a
function of $T$ (rather than $\chi _{spin}$). They observed a change of
slope around $240$ K and concluded the existence of a structural transition.
In fact, we find the $K$ vs $T$ data to be Curie-Weiss-like and the "change
of slope" seen in Ref. [\cite{AAGipplus1989}] is not found by us (our
temperature range is limited to 300 K, however). \ A progressive change in
the shape of the spectrum takes place below $200$ K and a large asymmetry is
seen. This change with $T$ may be related to structural distortions which
creates inequivalent sites. \ 

\paragraph{Spectra at low-T (below 80 K):}

The spectra broaden as $T$ decreases due to a distribution of shifts which
grows with decreasing temperature. \ Given the Curie-Weiss nature of the
spin susceptibility above $T_{N}$, a distribution of local environments for
the $^{45}$Sc nucleus will give this effect. \ In addition, presence an
anisotropy of the susceptibility will as well contribute to an increased
linewidth. \ \ Such broad spectra are measured more efficiently by sweeping
the field at a fixed frequency rather than the Fourier transform technique
used for $T\geq 80$ K$.$ We measured the spectra at different frequencies $%
25674$ kHz and $61674$ kHz in the $T-$range from $1.2$ to $80$ K and are
plotted in Fig. $11$ and $12,$ respectively. Fig. $11$ shows the measured
intensity of NMR spectra as a function of $T$ from $60$ K to $1.2$ K in an
applied field of about $25$ kOe. Since the spectra are measured in low
fields, the quadrupolar satellites are clearly seen at $60$ K compared to
the spectra measured in higher fields (see Fig. $12$).

We find that the spectrum broadens as the ordering temperature is approached
and the intensity is effectively lost near the ordering temperature at about 
$T=15$ K (see left inset of Fig. 11). The critical slowing down of
fluctuations near $T_{N}$ causes a divergence of the relaxation rate and an
effective disappearance of the signal. The spectrum reappears below the
ordering temperature and the intensity increases rapidly with decreasing $T$%
. The increase of the linewidth with decreasing temperature (in the ordered
state) indicates that $^{45}$Sc experiences a distribution of static local
fields ($^{45}$Sc is hyperfine coupled to inequivalent copper ions) which
gives the large broadening. For better visual clarity, we have shown the
normalized spectra in the $T-$range from $10$ K to $1.2$ K separately in the
inset of Fig. $11$.\ The full width at half maximum at the lowest
temperature measured ($1.2$ K) is about $3000$ Oe from the spectrum at $%
25674 $ kHz while the total spread of the line is close to 5000 Oe. \ This
is a measure of the field distribution seen at the $^{45}$Sc site. \ From
the crystal structure, it appears that $^{45}$Sc is hyperfine coupled to
three Cu ions of which two are equivalent. \ The total hyperfine coupling
was found to be -4.5 kOe. \ Therefore, the coupling per Cu is about -1.5 kOe/%
$\mu _{B}$. Therefore with a 1 $\mu _{B}$ magnetic moment localized at each
Cu site, one would expect the total spread of the line to be around 4.5 kOe
consistent with our observation. \ In typical cases where antiferromagnetic
ordering is observed, a single NMR line splits into two below the ordering
temperature. \ This happens due to the internal field (at a magnetic site)
either adding to or opposing the external magnetic field. \ However, one
needs single crystals in order to observe the abovementioned splitting. \ In
the present case, we have polycrystalline samples (efforts to align these
powders in an external field were not successful). \ Here, a rectangular
lineshape might be expected in the ordered state if the internal field at
the $^{45}$Sc sites is small compared to the applied field. \ \ Since there
are inequivalent $^{45}$Sc sites due to local distortions/defects as already
suggested by the room-temperature NMR spectrum in addition to a quadrupolar
broadening, \ it results in a generic dome-shaped lineshape.

\bigskip

In contrast to our $\mu $SR measurements which were done in zero field, our
NMR measurements have been done in various applied fields. \ Here, we are
able to track the changes in the magnetic state due to an applied field.
There are two main pieces of information that are obtained: (i) There is a
decrease of the ordering temperature due to an applied field. In Fig. $12$,
the measured NMR spectra at a fixed frequency at $61674$ kHz (i.e. $H=63$
kOe) in the $T-$range from $40$ K to $1.2$ K are shown. In this field, the
signal is retained down to $1.2$ K (see also the inset of Fig. $12$) due to
the absence of ordering. (ii) There is possibly a change in the
spin-ordering as a function of temperature in the ordered state. \ As we see
in the inset of Fig. 11, a broad feature (indicated by an arrow mark) is
present at $T=1.2$ K and it vanishes progressively as the temperature is
increased to $T=10$ K. In the variation of bulk magnetization with $H$ (see
Fig. $3$), there is a change in the slope of $M(H)$ at $23$ kOe, which is
smeared out in $M(H)$\ isotherms at high-$T$. So, the appearance of the
broad feature in the NMR spectrum might be associated with the spin-flopped
phase. \ In addition we have also plotted the spectra measured at $T=1.2$ K
at different frequencies $\nu _{ref}=18725$ kHz, $\nu _{ref}=25674$ kHz, $%
\nu _{ref}=29725$ kHz, and $\nu _{ref}=61674$ kHz in Fig. $13(a),(b),(c),$
and $(d)$, respectively. The line width at $1.2$ K is smallest for highest
frequency and/or field (where there is no ordering) and progressively
increases at lower fields where the internal field distribution affects
significantly the line shape. Particularly the line shape of the spectra
corresponding to $\nu _{ref}=18725$ kHz and $\nu _{ref}=25674$ kHz (which
are in magnetic fields below $25$ kOe) are different from the shape of the
spectrum measured at $\nu _{ref}=29725$ kHz. This means that internal field
below $25$ kOe (where a spin-flop transition is observed) is different from
the internal fields above $30$ kOe. This change might be associated with the
spin-flop transition observed in $M(H)$.

\bigskip

Thus, the information obtained from the local probe measurements mentioned
above could be summarized as follows. \ Firstly, the existence of LRO at 16
K from the bulk of the sample is confirmed by $\mu $SR in zero field. \ The
linear scaling of the NMR shift with the measured susceptibility ensures
that the $T$-dependent part of the measured susceptibility is in fact the
intrinsic spin susceptibility confirming the sample to be free from magnetic
defects/impurities. \ The loss in the intensity of the NMR signal well above
the ordering temperature is indicative of the presence of short-range order
or a large $T$-regime of slow fluctuations. \ The field dependence of the
ordering is also exhibited in the NMR measurements. \ In order to throw
light on the various magnetic interactions which are important in BCSO, it
is necessary to make an effort on the theoretical front. \ This is carried
out in the next section following which we arrive at a fuller understanding
of BCSO.

\bigskip

\subsection{Electronic structure calculations}

Our analysis of the electronic structure of Ba$_{3}$Cu$_{3}$Sc$_{4}$O$_{12}$
are carried out in the framework of the tight-binding linearized muffin tin
orbital (TB-LMTO) method in the atomic sphere approximation (ASA) within
local density approximation to the density functional theory\cite{tblmto47}.
The space filling in the ASA is achieved by inserting appropriate empty
spheres and the TB-LMTO ASA band structure results are found to be in good
agreement with the full potential linearized augmented plane wave (FP-LAPW)
calculations as implemented in WIEN-2K code\cite{wien2K}. The basis set for
the self-consistent electronic structure calculations for Ba$_{3}$Cu$_{3}$Sc$%
_{4}$O$_{12}$ in TB-LMTO ASA includes Ba (\textit{s},\textit{d},\textit{f})
, Cu (\textit{s},\textit{p},\textit{d}), Sc(\textit{s},\textit{d}) and O(%
\textit{s},\textit{p}) and the rest are downfolded. A (16,16,16) \textit{k}%
-mesh has been used for self-consistency. All the \textit{k}-space
integrations were performed using the tetrahedron method\cite{tetrahedron}.
In order to extract a low energy tight-binding (TB) model Hamiltonian and
its various hoppings we have employed the N-th order muffin tin orbital
(NMTO) based downfolding method \cite{nmto}. For the calculations of the
various exchange interactions we have performed total energy calculations in
the framework of GGA+U method for various ordered spin states. In order to
extract the various exchange interactions, the relative energies of these
ordered spin states determined from the GGA+U calculations are mapped onto
the corresponding energies obtained from the total spin exchange energies of
the Heisenberg spin Hamiltonian $H=-\sum_{i,j}J_{ij}\vec{S}_{i}\cdot \vec{S}%
_{j}$. The total energy calculations are carried out in the plane wave basis
along with the projector augmented wave (PAW) method \cite{paw} as
implemented in Vienna Ab-Initio Simulation Package (VASP) \cite{vasp, vasp2}%
. The energy cut-off for the plane wave expansion of the PAW's was taken to
be 500 eV. The GGA+U calculations \cite{dudarev} are carried out for three
different values of on-site $d$-$d$ Coulomb interaction for Cu \emph{viz.} U
= 4 eV, 6 eV, and 8 eV while the onsite exchange parameter was fixed at 1 eV.

The all orbital band structure along the various high symmetry points of the
body centered tetragonal BZ obtained by TB-LMTO-ASA calculations is
displayed in Fig. 14. All the energies are measured with respect to the
Fermi level $E_{f}$. The characteristic feature of the band structure are
half-filled six Cu-$d$ bands near the Fermi level. These bands arise from
the six Cu atoms in the unit cell and are predominantly Cu \textit{d}$%
_{x^{2}-y^{2}}$ character in a local frame of reference where Cu is at the
origin and the $x$ and $y$ axes point along the oxygens residing on the same
square plaquette. This six band complex is separated from the other
completely filled non Cu- \textit{d}$_{x^{2}-y^{2}}$ bands and O p bands by
a gap of about 0.9 eV. The six band complex hybridizes strongly with O1 and
O3 forming a square plaquette for Cu1 and Cu2 respectively and hardly with
O2 which does not reside on these square plaquettes. This isolated Cu-d$%
_{x^{2}-y^{2}}$ six band complex is responsible for the low energy physics
of this material. While the system is metallic in LDA (see the DOS in the
inset), inclusion of Coulomb correlations, within LSDA+U makes the system
insulating.

As the six band manifold is well separated from the rest, so a tight-binding
model with a single Cu \textit{d}$_{x^{2}-y^{2}}$ orbital per Cu site would
be a good approximation to the full band structure close to the Fermi level.
We have employed NMTO downfolding method to construct a low energy, few band
tight-binding model Hamiltonian for this system. The down-folding method
consists of deriving a few-orbital effective Hamiltonian from the
all-orbital LDA Hamiltonian by downfolding the inactive orbitals into the
tails of the active orbitals kept in the basis chosen to describe the low
energy physics. This method results in renormalized effective interactions
between the active orbitals (here Cu-\textit{d}$_{x^{2}-y^{2}}$) retained in
the basis. In the present case the low-energy bands form an isolated set of
bands and the constructed effective orbitals, the NMTO's are the Wannier
functions corresponding to the low energy bands. A real space representation
of the downfolded Hamiltonain via Fourier \ transform $\mathcal{H}(\mathbf{k}%
)$ $\rightarrow $ $\mathcal{H}(\mathbf{R})$ $\left( \mathcal{H}(\mathbf{R}%
)=\sum_{\langle i,j\rangle }t_{ij}\left( c_{j}^{\dagger
}c_{i}+c_{i}^{\dagger }c_{j}\right) \right) $ yields the effective hopping
parameters between the Cu ions. These hoppings will determine the dominant
exchange paths and in turn will clarify the magnetic dimensionality of the
system. The downfolded Cu-\textit{d}$_{x^{2}-y^{2}}$ bands are plotted in
Fig. 14 and we note that the agreement with the full band structure is
remarkable thereby justifying our low energy model Hamiltonian. Table $2$
shows the various dominant effective hopping integrals between Cu ions at
site $i$ and $j$. \ These have been abbreviated as $t_{n}$ where $n$ is a
number indicating the path shown in Fig. 1 We gather from Table $2,$ that
the strongest hopping is $t_{2}$ between Cu2 and Cu2 in a particular
diamond. As anticipated, Cu2 and Cu1 hopping in the diamond is weak due to
its geometry, while unexpectedly Cu1 - Cu1 coupling along the chain is quite
strong. In order to obtain insights on the Cu1 - Cu1 hopping $t_{3}$ we have
plotted the Cu1 Wannier function in Fig. 15. We find from the figure that
Cu1-\textit{d}$_{x^{2}-y^{2}}$ strongly antibinds with the neighboring
oxygens which in turn antibind with Cu2 and again via oxygen to Cu1. It is
interesting to note that the oxygen tails bend in order to strengthen the
Cu-O bonding and therefore the Cu1-O-Cu2-O-Cu1 super-super-exchange path
(SSE) resulting in a dominant $t_{3}$ interaction which is otherwise
prohibited by symmetry. Table $2$ also reveals that in addition to the
intra-chain hoppings the inter-chain hoppings are quite dominant .

Guided by the various hopping strengths we have computed twelve exchange
interactions for three different U values and the results of our
calculations are presented in Table $2$. The antiferromagnetic contribution
to the exchange using second-order perturbation is given by $J_{i}^{AFM}=-%
\frac{4t_{i}^{2}}{U_{eff}}$ and is also listed in Table $2$; here $t_{i}$
are various hoppings and $U_{eff}$ is the effective Coulomb interaction
taken to be 4 eV in our calculations. From Table 2 we gather that the
dominant exchange interaction $J_{1}$ is ferromagnetic, while $J_{2}$ , $%
J_{3}$, and $J_{4}$ are antiferromagnetic. As expected, antiferromagnetic $%
J_{2}$, $J_{3}$ and $J_{4}$ decrease with increasing $U$ values. Further,
inter-chain interactions $J_{5}$, $J_{9}$ and $J_{11}$ are ferromagnetic. We
note both intra-chain as well as inter-chain interactions are substantial
accounting for the magnetic long range order seen for this system. The high
temperature behaviour is dominated by the strong nearest-neighbor
ferromagnetic exchange interaction $J_{1}$. This strong ferromagnetic
exchange follows from the Anderson-Goodenough-Kanamori rules \cite%
{Goodenough, Kanamori} as the $\angle \mathrm{Cu1}-\mathrm{O1}-\mathrm{Cu2}$
is 86.84$^{\circ }$ (close to 90$^{\circ }$). At low temperature further
long range interaction dominates primarily mediated by oxygen via
super-super exchange and accounts for the antiferromagnetic behaviour. An
estimate of the asymptotic Curie-Weiss temperature $\theta $ obtained from
the calculated exchange interactions within the mean field approximation 
\cite{whangbo} yields $\theta $ = 96 K, 68 K and 55 K for $U$ =4, 6, 8 eV
respectively, in good agreement with our experimental result (65 K).

\section{Summary and Conclusion}

Our work on Ba$_{3}$Cu$_{3}$Sc$_{4}$O$_{12}$ was motivated by its crystal
structure which suggested the possibility of exotic magnetic behaviour due
to the apparent diamond-chain-like configuration of the magnetic Cu atoms. \
Whereas our detailed work has demonstrated the dominance of 3D magnetic
interactions, there are several interesting features in the magnetism of
this compound. \ The high-$T$ magnetic susceptibility is Curie-Weiss-like
with a positive asymptotic Weiss temperature suggestive of \textit{%
ferromagnetic} interactions. \ In contrast, the ordering is found to be
antiferromagnetic with the magnetic moments at the Cu1 and Cu2 sites are
aligned opposite to each other. \ This has to be seen in light of our
electronic structure calculations which indicate comparable Cu1-Cu1 and
Cu2-Cu2 interactions. \ A possible scenario that we have discussed in this
paper is that the stronger Cu1-Cu2 ferromagnetic interaction (this exchange
coupling is of the order of 160 K) drives the high-$T$ Curie-Weiss behaviour
while the weaker Cu1-Cu1 and the Cu2-Cu2 antiferromagnetic exchanges (of the
order of 50 K) drive the transition to AF order at low-$T$. \ The
sensitivity of the transition temperature to an applied magnetic field (a
field of 70 kOe was sufficient to decrease the $T_{\text{N}}$ from $16$ K to
zero while one might have naively expected this to happen in a field of
about 240 kOe) is, at first sight, somewhat puzzling and indicates the
presence of competing interactions. \ This is also suggested by the
heat-capacity measurements where the observed entropy change due to the AF
ordering is found to be smaller than the expected value of $R\ln $2. \
Another interesting aspect is the change in slope in the low-temperature $M$
vs $H$ isotherms. \ This, coupled with the observed difference in the $^{45}$%
Sc NMR lineshapes at 1.2 K in various applied fields points towards a change
in the spin ordering with increasing applied field, at around 23 kOe. \ Our
detailed electronic structure calculations have indicated the presence of a
dominant ferromagnetic interaction as also several, somewhat weaker
antiferromagnetic interactions. \ These calculations then enable us to
qualitatively reconcile with our experimental data. \ We note that our $\mu $%
SR measurements were done in zero field and the absence of \ any change in
spin ordering seen there is not in contradiction with the above data. \ A
yet another interesting feature has to do with the crystal structure. \
Since the $^{45}$Sc nucleus has $I>1/2$ its NMR spectrum should be a
quadrupole powder pattern provided the site symmetry of the $^{45}$Sc
nucleus is lower than cubic. \ In Ba$_{3}$Cu$_{3}$Sc$_{4}$O$_{12}$, there
are ScO$_{6}$ octahedra with only a slight deviation from the perfect
octahedral environment (i.e., only a small crystalline electric field
gradient \textit{efg} at the center of each octahedron). \ Accordingly, the
room temperature NMR spectrum of $^{45}$Sc in Ba$_{3}$Cu$_{3}$Sc$_{4}$O$%
_{12} $ exhibits weak satellite intensities suggesting that many of the
sites feel a negligible \textit{efg} while the remaining give rise to a
quadrupole-broadened powder pattern, presumably due to structural
distortions around them. \ 

To conclude,\ our extensive measurements coupled with \ \textit{ab initio}
electronic structure calculations (where we have determined the relative
strengths and the signs of the exchange interactions between various copper
ions and also the asymptotic Weiss temperature $\theta $) have helped in
developing a comprehensive understanding of the magnetic properties of Ba$%
_{3}$Cu$_{3}$Sc$_{4}$O$_{12}$. \ In particular, we find that our work shows
that the system is complex and a naive assessment of the crystal
structure/atomic arrangement to arrive at a conclusion regarding the
magnetic dimensionality can be misleading and a theoretical determination of
the various important magnetic couplings is essential for a complete picture.

\begin{acknowledgments}
We thank the Indo-French Center for the Promotion of Advanced Research and
ARCUS Ile de France-Inde. SR and VS are grateful to Prof. A.K. Raychaudhuri
and DST, India for travel funding. \ JC acknowledges CSIR, India for a
research fellowship. We thank A. Amato for technical support with the GPS
instrument at the Swiss Muon Source, Paul Scherrer Institute, Villigen,
Switzerland where the $\mu $SR work was peformed.
\end{acknowledgments}

\textbf{Table Caption}

Table 1: Structural parameters after the Rietveld refinement of
neutron-diffraction pattern of Ba$_{3}$Cu$_{3}$Sc$_{4}$O$_{12}$ at 300 K and
2 K.

%\begin{ruledtabular}
\begin{tabular}{|l|l|l|}
\hline
Parameters & $T=300$ K & $T=2$ K \\ \hline
space group & \textit{I4/mcm} & \textit{I4/mcm} \\ \hline
a & 11.916(4) \AA  & 11.913(4) \AA  \\ \hline
c & 8.406(4) \AA  & 8.404(4) \AA  \\ \hline
volume & 1193.83(2) \AA $^{3}$ & 1192.82(2) \AA $^{3}$ \\ \hline
overall isotropic displacement factor, Bov & 1.95 \AA $^{2}$ & 0.225 \AA $%
^{2}$ \\ \hline
$\chi ^{2}$ & 4.10 & 4.77 \\ \hline
R$_{p}$ (\%) & 2.52 & 1.35 \\ \hline
R$_{wp}$ (\%) & 3.26 & 1.76 \\ \hline
R$_{exp}$ (\%) & 1.61 & 0.81 \\ \hline
$\mu _{Cu1}(\mu _{B})$ &  & -0.65 \\ \hline
$\mu _{Cu2}(\mu _{B})$ &  & 1.15 \\ \hline
\end{tabular}%
%
%
%
%
%
%
%
%
%
%
%
%
%
%
%
%
%
%
%
%
%
%
%
%
%
%
%
%
%
%
%
%
%
%
%
%
%
%
%
%
%
%
%
%
%
%
%
%
%\end{ruledtabular}

Table 2 Hopping integrals ($t_{n}$) and exchange interactions ($J$) in meV.

%\begin{ruledtabular}
\begin{tabular}{|c|l|c|c|c|c|}
\hline
Cu-Cu distance & Hopping & $J_{i}^{AFM}=-4t_{i}^{2}/U_{eff}$ & Exchange & 
Exchange & Exchange \\ 
(\AA ) & (meV) & (meV) & (meV) & (meV) & (meV) \\ 
\  &  &  & $U=4$eV & $U=6$eV & $U=8$eV \\ \hline
2.74 & $t_{1}=$16.3 & -0.27 & 14.58 & 12.40 & 13.88 \\ 
3.53 & $t_{2}=$53.0 & -2.81 & -8.22 & -8.23 & -6.93 \\ 
4.19 & $t_{3}=$45.0 & -2.02 & -5.58 & -3.23 & -2.74 \\ 
4.88 & $t_{4}=$13.6 & -0.18 & -10.5 & -8.13 & -2.53 \\ 
6.88 & $t_{5}=$8.1 & -0.07 & 9.06 & 7.11 & 1.56 \\ 
6.97 & $t_{6}=$5.4 & -0.03 & -0.21 & -0.09 & -0.05 \\ 
8.39 & $t_{7a}=$8.0 & -0.06 & 0.32 & 0.38 & 1.03 \\ 
8.39 & $t_{7b}=$13.6 & -0.19 & 0.63 & 0.76 & 2.06 \\ 
8.41 & $t_{8}=$19.0 & -0.36 & 1.70 & -1.24 & -3.89 \\ 
9.15 & $t_{9}=$14.9 & -0.22 & 7.51 & 5.70 & -0.03 \\ 
9.40 & $t_{10}=$18.0 & -0.32 & -2.89 & -2.21 & -1.30 \\ 
9.73 & $t_{11}=$16.3 & -0.27 & 3.28 & 3.32 & -0.22 \\ \hline
\end{tabular}%
%
%
%
%
%
%
%
%
%
%
%
%
%
%
%
%
%
%
%
%
%
%
%
%
%
%
%
%
%
%
%
%
%
%
%
%
%
%
%
%
%
%
%\end{ruledtabular}

%Table 1 Hopping integrals ($t_{n}$) in meV obtained from NMTO\ downfolding
%method.

%\begin{tabular}{|c|c|c|}
%\hline
%Hopping path & Cu-Cu distance & $t_{n}$ \\ \hline
%& in \AA  & in meV \\ \hline
%Cu1-Cu2 ($t_{1}$) & 2.742 & 16.3 \\ \hline
%Cu2-Cu2 ($t_{2}$) & 3.527 & 53 \\ \hline
%Cu1-Cu1 ($t_{3}$) & 4.197 & 45 \\ \hline
%Cu2-Cu2 ($t_{4}$) & 4.882 & 13.6 \\ \hline
%Cu1-Cu2 ($t_{5}$) & 6.88 & 13.6 \\ \hline
%Cu2-Cu2 ($t_{6}$) & 6.974 & 5.4 \\ \hline
%Cu1-Cu2 ($t_{7}$) & 8.394 & 8.1 \\ \hline
%Cu2-Cu2 ($t_{8}$) & 8.414 & 19 \\ \hline
%Cu1-Cu2 ($t_{9}$) & 9.403 & 18 \\ \hline
%\end{tabular}

\textbf{Figure Captions}

Figure 1: The crystal structure of Ba$_{3}$Cu$_{3}$Sc$_{4}$O$_{12}$. The
diamond like\ chains formed by Cu1 and Cu2 atoms, are travelling along the $%
c $-direction. \ The notation for the various hopping integrals is also
indicated.

Figure 2: (a) The $T$-dependence of magnetic susceptibility (left $y-$axis)
and inverse susceptibility (right $y-$axis)\ of Ba$_{3}$Cu$_{3}$Sc$_{4}$O$%
_{12}$ are plotted. (b) $\chi (T)$in various applied fields. The ordering is
seen to be suppressed by the applied fields.

Figure 3: Magnetization as a function of the applied field $M$$(H)$ measured
at various temperatures from 2 K to 20 K. Inset shows the plot of $M(H)$ at
2 K on the left $y$-axis. The derivative of the magnetization is plotted on
the right $y$-axis. The solid line is the generated Brillouin function for $%
S=%
%TCIMACRO{\U{bd}}%
%BeginExpansion
{\frac12}%
%EndExpansion
$ and $g=2.1$ at $T=2$K.

Figure 4: (a) Heat-capacity vs temperature with a fit (explained in the
text). Inset shows the plot of the magnetic heat-capacity $C_{M}$ (on left $%
y-$axis) and magnetic entropy $S_{M}$ (on right $y-$axis) versus $T$ (b)
Variation of heat-capacity divided by temperature $(C_{p}/T)$ as a function
of $T$ in various applied fields from $0$ to $9$ Tesla.

Figure 5: The variation of $T_{N}$ with applied fields. The data points in
the plot separate the antiferromagnetic (AF) and paramagnetic (PM) regions.

Figure 6 (a): Neutron Diffraction (ND) pattern for Ba$_{3}$Cu$_{3}$Sc$_{4}$O$%
_{12}$ measured at $T$ = 300 K refined using Rietveld refinement method. The
vertical tick marks indicate the nuclear Bragg peak positions. (b) Rietveld
refined low temperature ND patterns for Ba$_{3}$Cu$_{3}$Sc$_{4}$O$_{12}$.
The magnetic peak (0 0 0) is marked for $T$ = 2 K. The nuclear and magnetic
Bragg peak positions are indicated by N and M respectively. The inset (i)
shows the decrease in the integrated intensity of the magnetic peak with
increasing temperature. The inset (ii) shows the variation of magnetic
moments of Cu1 and Cu2 with temperatures.

Figure 7: ND patterns for Ba$_{3}$Cu$_{3}$Sc$_{4}$O$_{12}$ measured at $T$ =
2 K and various applied magnetic fields. The magnetic peak (0 0 0) broadens
with increasing magnetic field.

Figure 8: Top panel: Zero field asymmetry at 1.5 K. In the expanded view of
the inset, one can clearly identify the beats between several oscillating
components. Bottom panel: Evolution of the asymmetry around the transition
from a slowly relaxing Gaussian shape at 16.58~K to an oscillating pattern
already visible at 16.19~K.

Figure 9: Left panel: Fourier transform of the oscillating asymmetry. Four
muon sites are clearly evidenced. Right panel: Temperature evolution of the
frequencies corresponding to the four oscillating components of the muon
asymmetry. In the lower graph, the variation of the normalized frequencies
is plotted. The the solid line is a fit to $(1-(T/T_{N})^{\beta }$ (see
text).

Figure 10: Normalized $^{45}$Sc NMR spectra vs. normalized frequency $(\nu
-\nu _{ref})/\nu _{ref}$, where $\nu _{ref}=(\gamma /2\pi )H$ is the
reference frequency. Inset (a) shows the measured NMR spectrum as a function
of frequency ($\nu $) which can be simulated by combining a quadrupole split
powder pattern (spectrum A) and a single line (spectrum B). The ratio of
spectrum A and B is 1:1. (b) shows NMR shift $K$ vs the spin susceptibility
along with a linear fit.

Figure 11: Measured $^{45}$Sc NMR spin-echo intensity as a function of the
applied magnetic field (swept-field experiment) at a fixed frequency $\nu
_{ref}=25674$ kHz for various temperatures from $60$ K to $12$ K. The left
inset shows the decrease of the NMR spectral intensity multiplied by
temperature as one approaches the ordering temperature. The right inset
shows the normalized NMR spectra for temperatures from $10$ K to $1.2$ K. \ 

Figure 12: Measured spin-echo intensities of $^{45}$Sc NMR line as a
function of the applied magnetic field (swept-field experiment) at a fixed
frequency $\nu _{ref}=61674$ kHz for various temperatures from $40$ K to $15$
K. Inset shows the NMR spectra for temperatures from $10$ K to $1.2$ K.

Figure 13: Variation of the $^{45}$Sc NMR spectra at various radio
frequencies at $T=1.2$ K. The line is seen to get broader for lower
frequencies (lower fields) due to the spin ordering related field
distribution at $^{45}$Sc site. The solid lines are the reference lines ($%
H_{ref}=2\pi \nu _{ref}/\gamma $) for the corresponding reference
frequencies.

Figure 14: Downfolded band structure (shown as a red line) compared with all
orbital LDA band structure. Inset shows the total density of states for Ba$%
_{3}$Cu$_{3}$Sc$_{4}$O$_{12}$.

Figure 15: Effective Cu1 $d_{x^{2}-y^{2}}$ Wannier function plot for Ba$_{3}$%
Cu$_{3}$Sc$_{4}$O$_{12}$.

\bigskip 

\FRAME{ftbpF}{3.8285in}{6.4022in}{0pt}{}{}{figure1.png}{\special{language
"Scientific Word";type "GRAPHIC";maintain-aspect-ratio TRUE;display
"ICON";valid_file "F";width 3.8285in;height 6.4022in;depth
0pt;original-width 3.7801in;original-height 6.3399in;cropleft "0";croptop
"1";cropright "1";cropbottom "0";filename '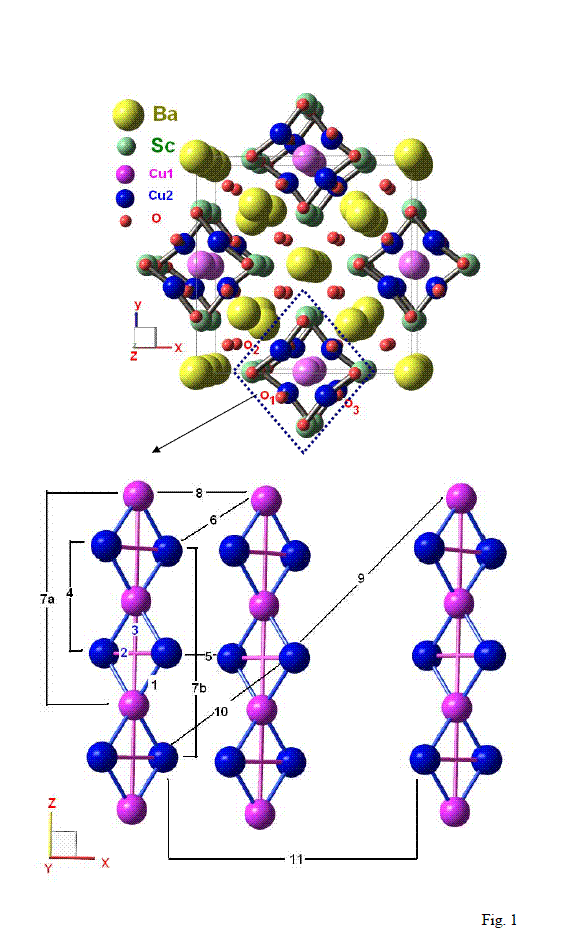';file-properties
"XNPEU";}}

\FRAME{ftbpF}{4.0395in}{5.22in}{0pt}{}{}{figure2.png}{\special{language
"Scientific Word";type "GRAPHIC";maintain-aspect-ratio TRUE;display
"ICON";valid_file "F";width 4.0395in;height 5.22in;depth 0pt;original-width
16.9668in;original-height 21.9533in;cropleft "0";croptop "1";cropright
"1";cropbottom "0";filename '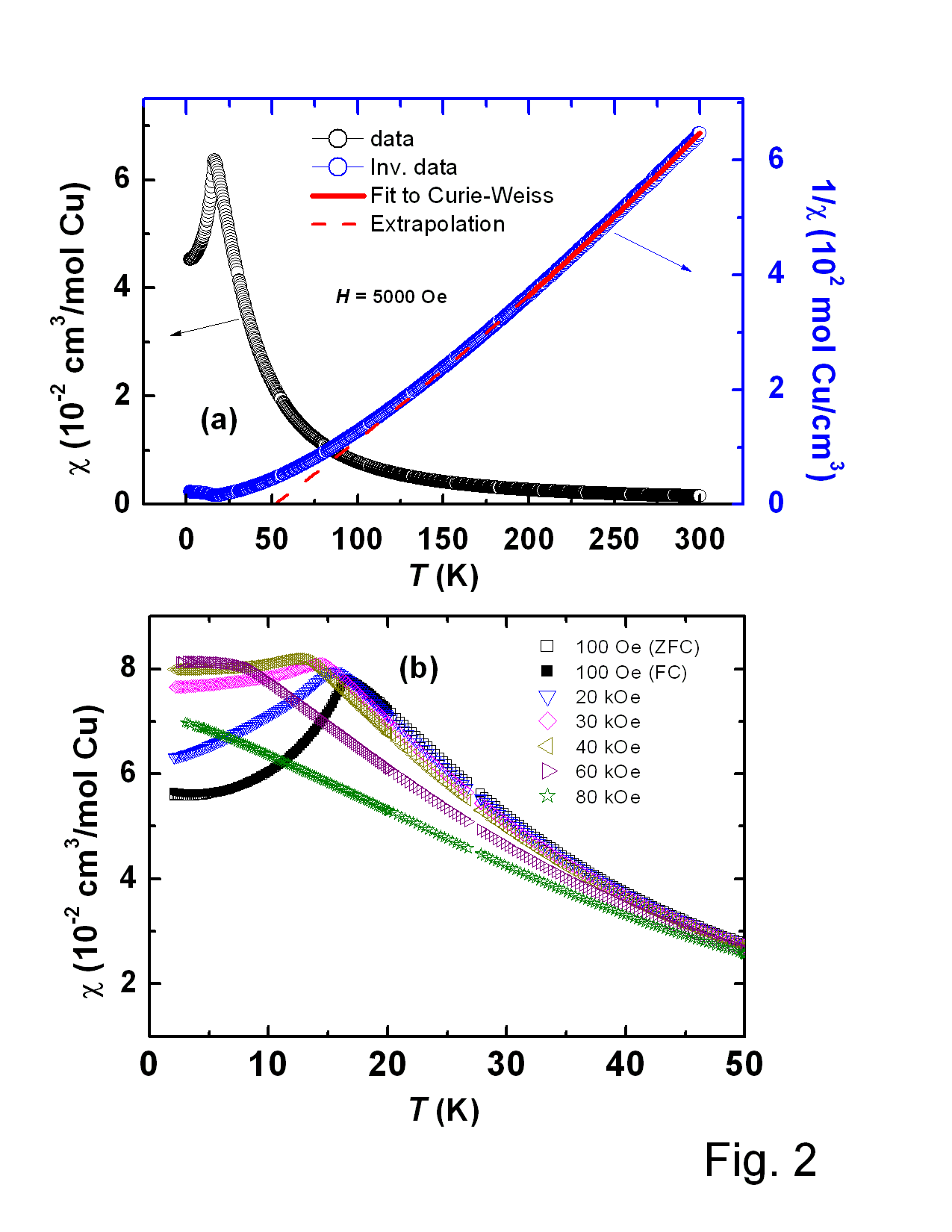';file-properties "XNPEU";}}

\FRAME{ftbpF}{6.7966in}{5.2209in}{0pt}{}{}{figure3.png}{\special{language
"Scientific Word";type "GRAPHIC";maintain-aspect-ratio TRUE;display
"ICON";valid_file "F";width 6.7966in;height 5.2209in;depth
0pt;original-width 20.8134in;original-height 15.967in;cropleft "0";croptop
"1";cropright "1";cropbottom "0";filename '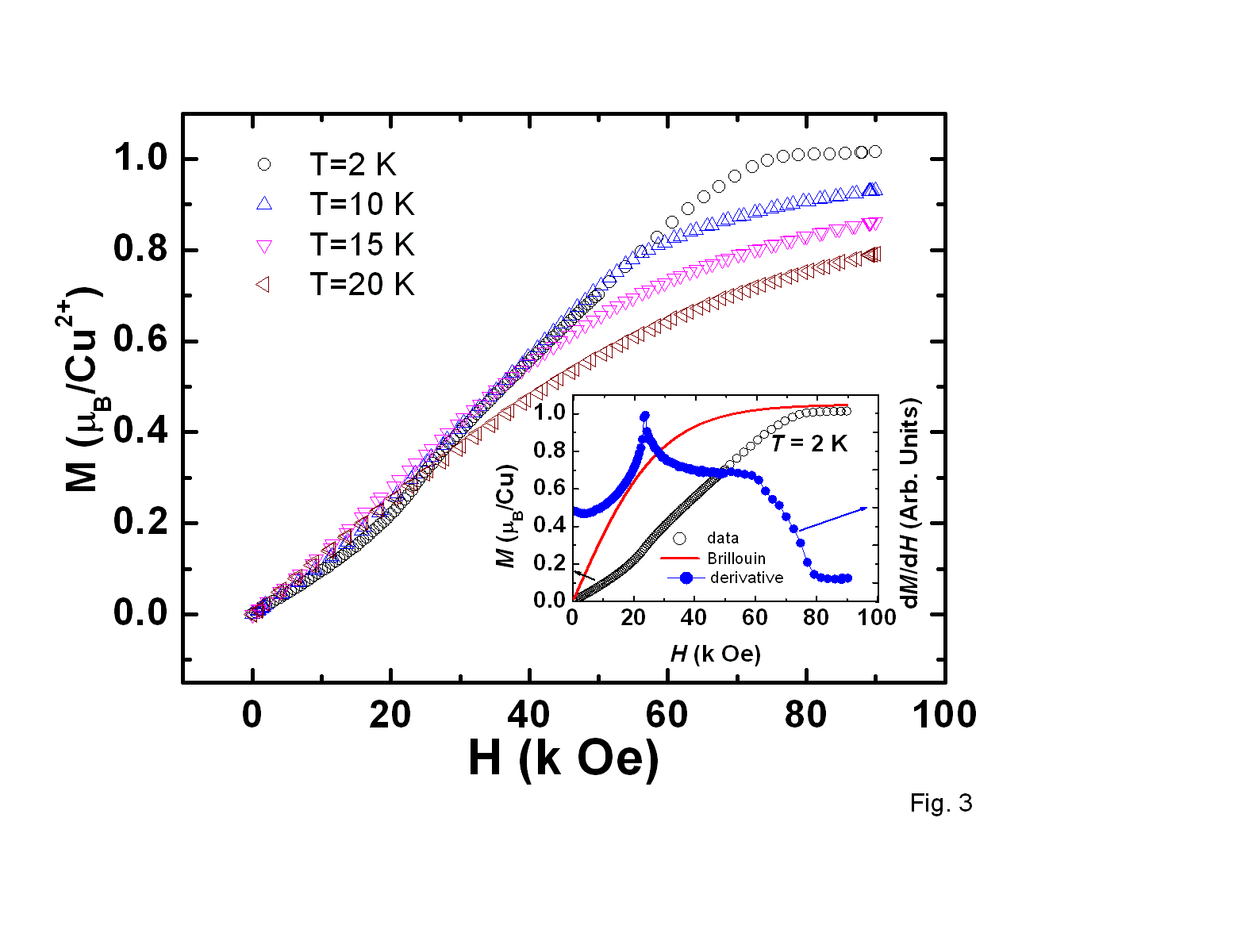';file-properties
"XNPEU";}}

\FRAME{ftbpF}{4.0395in}{5.22in}{0pt}{}{}{figure4.png}{\special{language
"Scientific Word";type "GRAPHIC";maintain-aspect-ratio TRUE;display
"ICON";valid_file "F";width 4.0395in;height 5.22in;depth 0pt;original-width
16.9668in;original-height 21.9533in;cropleft "0";croptop "1";cropright
"1";cropbottom "0";filename '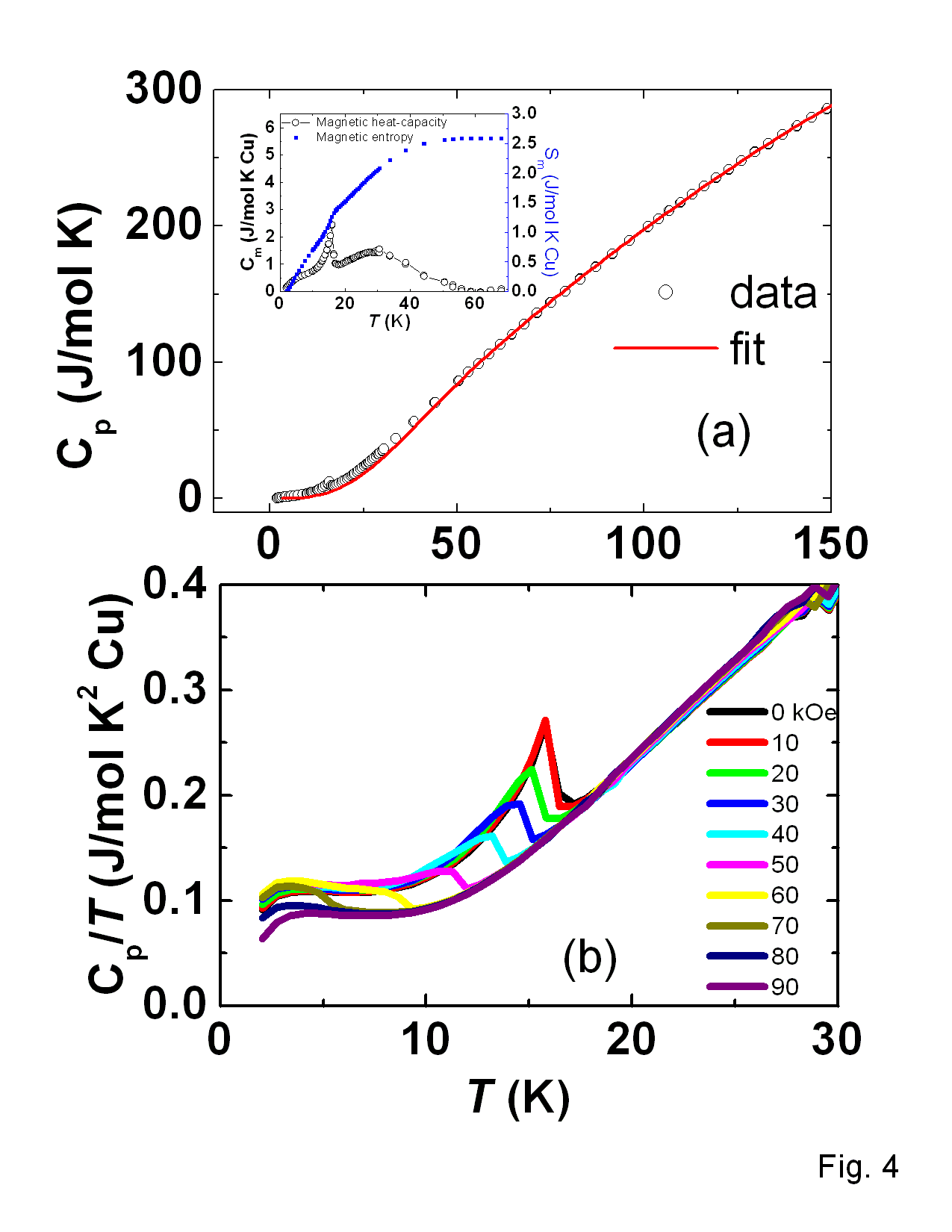';file-properties "XNPEU";}}

\FRAME{ftbpF}{4.0395in}{5.22in}{0pt}{}{}{figure5.png}{\special{language
"Scientific Word";type "GRAPHIC";maintain-aspect-ratio TRUE;display
"ICON";valid_file "F";width 4.0395in;height 5.22in;depth 0pt;original-width
16.9668in;original-height 21.9533in;cropleft "0";croptop "1";cropright
"1";cropbottom "0";filename '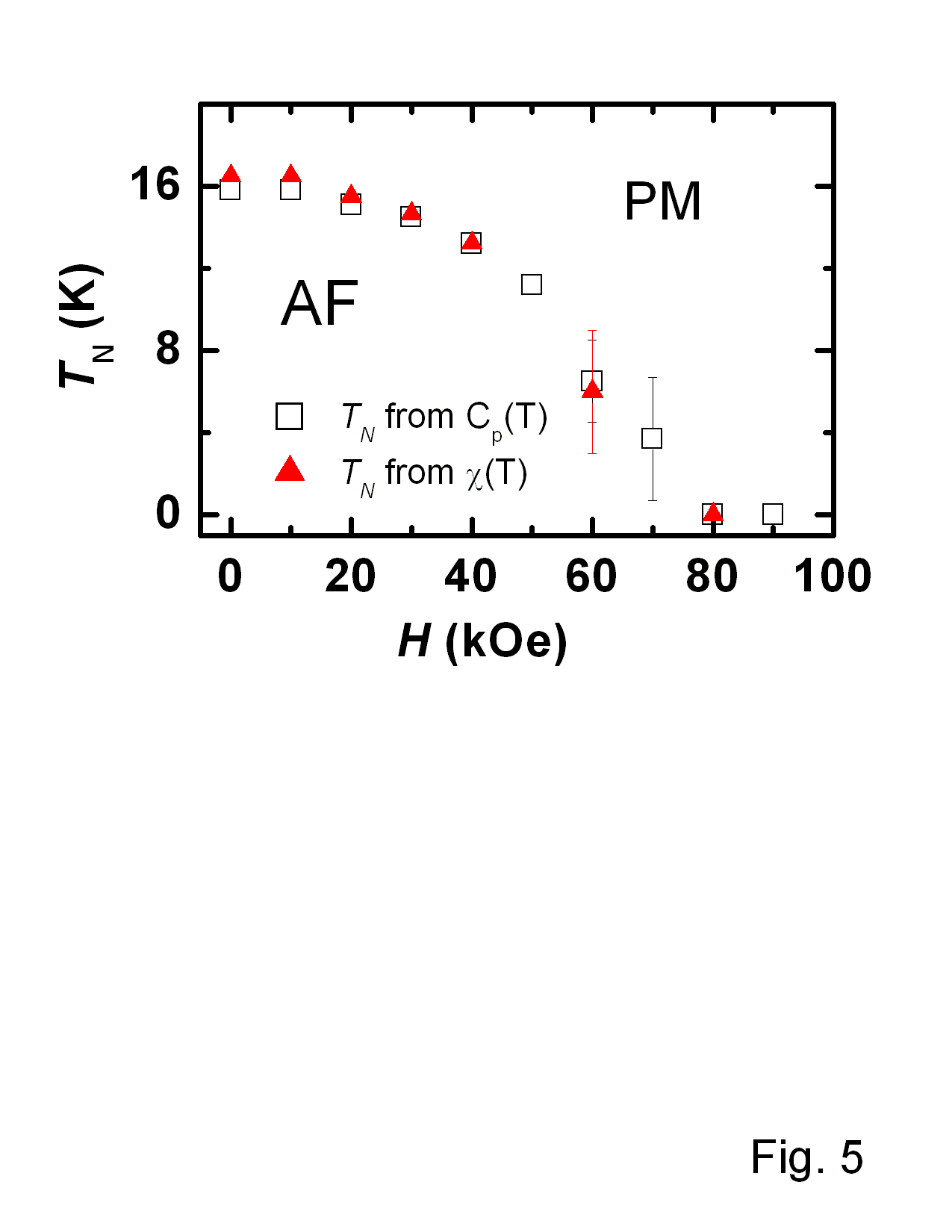';file-properties "XNPEU";}}

\FRAME{ftbpF}{3.6979in}{5.2209in}{0pt}{}{}{figure6.png}{\special{language
"Scientific Word";type "GRAPHIC";maintain-aspect-ratio TRUE;display
"ICON";valid_file "F";width 3.6979in;height 5.2209in;depth
0pt;original-width 16.4868in;original-height 23.3266in;cropleft "0";croptop
"1";cropright "1";cropbottom "0";filename '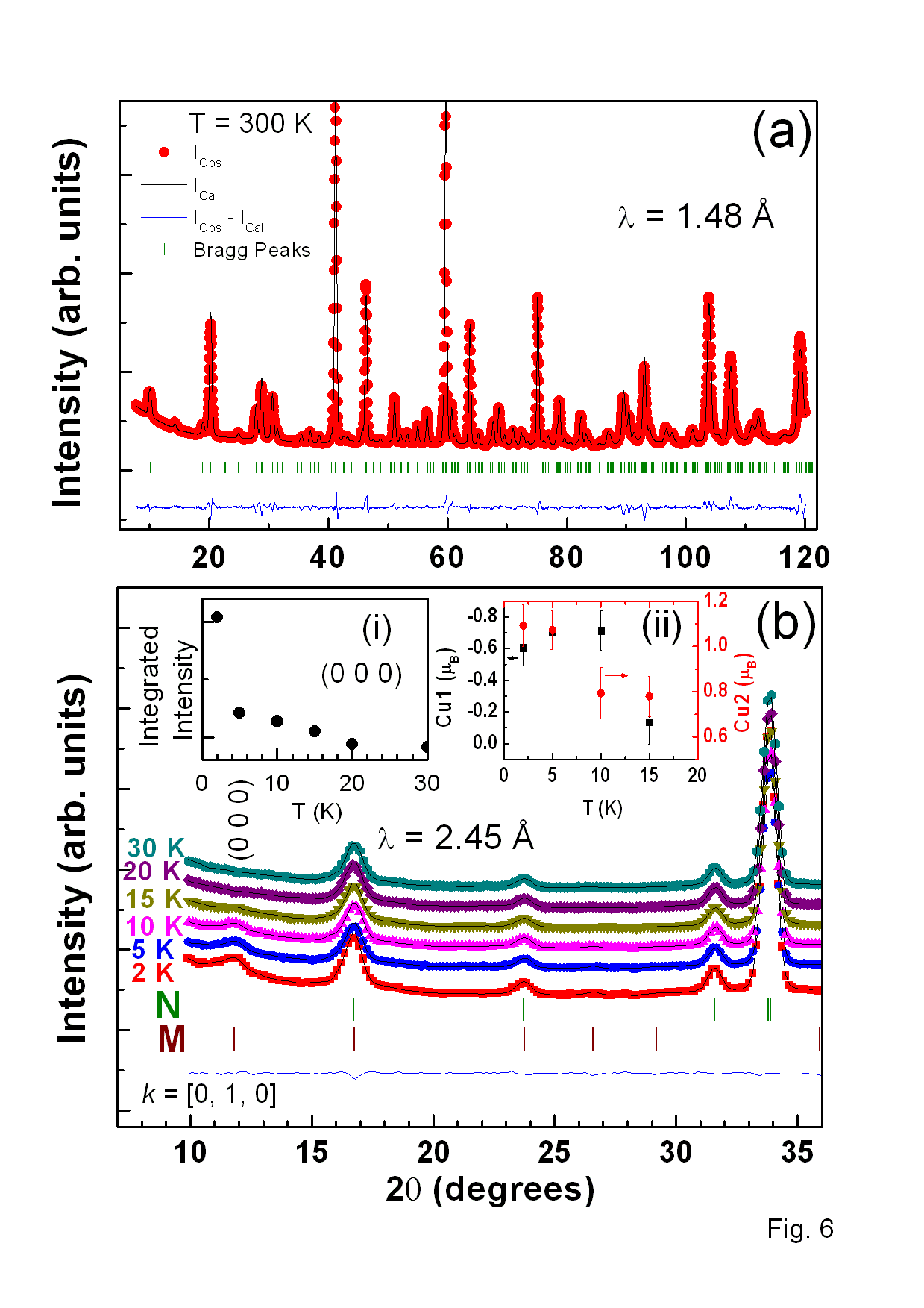';file-properties
"XNPEU";}}

\FRAME{ftbpF}{3.6979in}{5.2209in}{0pt}{}{}{figure7.png}{\special{language
"Scientific Word";type "GRAPHIC";maintain-aspect-ratio TRUE;display
"ICON";valid_file "F";width 3.6979in;height 5.2209in;depth
0pt;original-width 16.4868in;original-height 23.3266in;cropleft "0";croptop
"1";cropright "1";cropbottom "0";filename '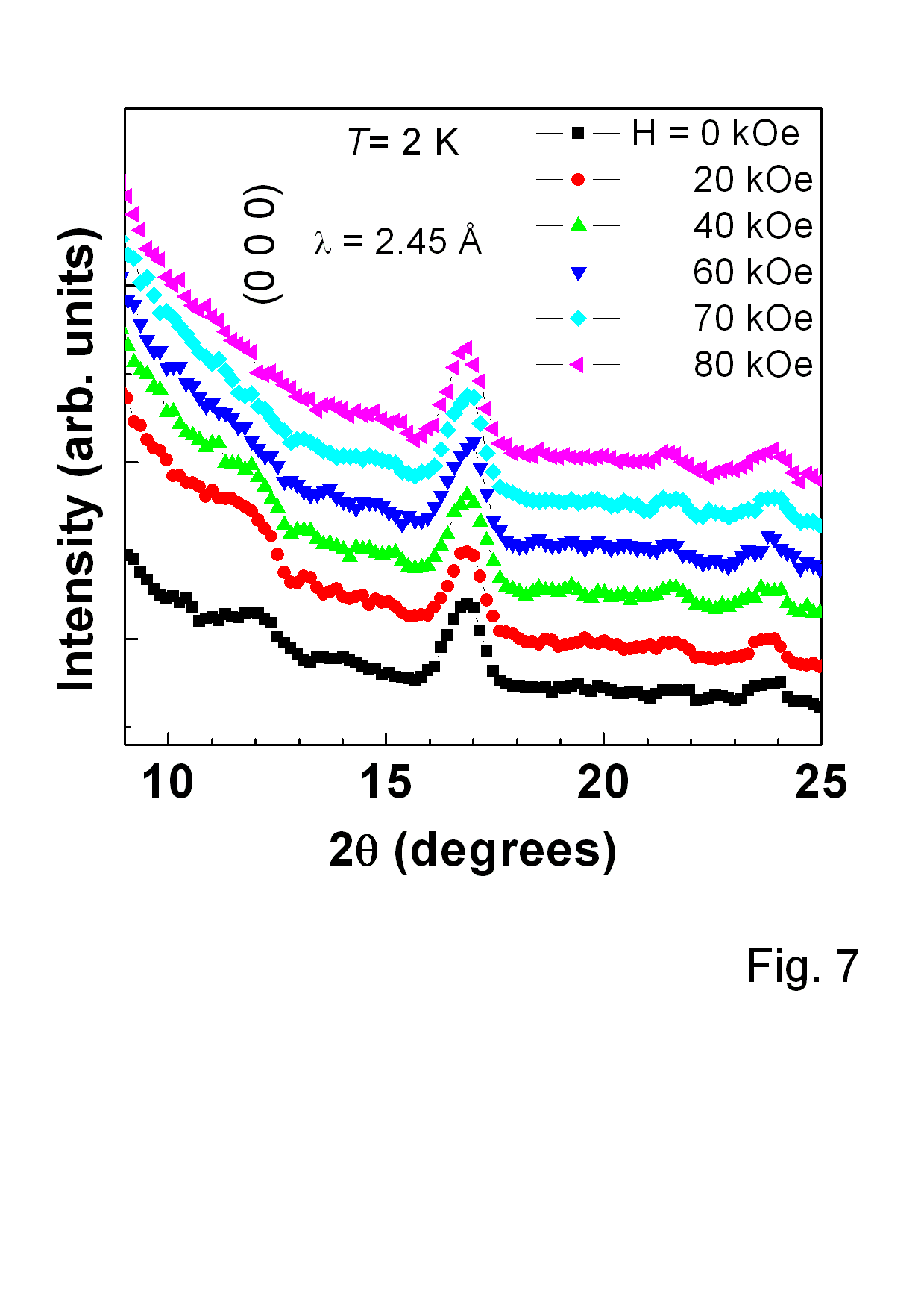';file-properties
"XNPEU";}}

\FRAME{ftbpF}{6.5207in}{5.2226in}{0pt}{}{}{figure8.png}{\special{language
"Scientific Word";type "GRAPHIC";maintain-aspect-ratio TRUE;display
"ICON";valid_file "F";width 6.5207in;height 5.2226in;depth
0pt;original-width 8.5331in;original-height 6.826in;cropleft "0";croptop
"1";cropright "1";cropbottom "0";filename '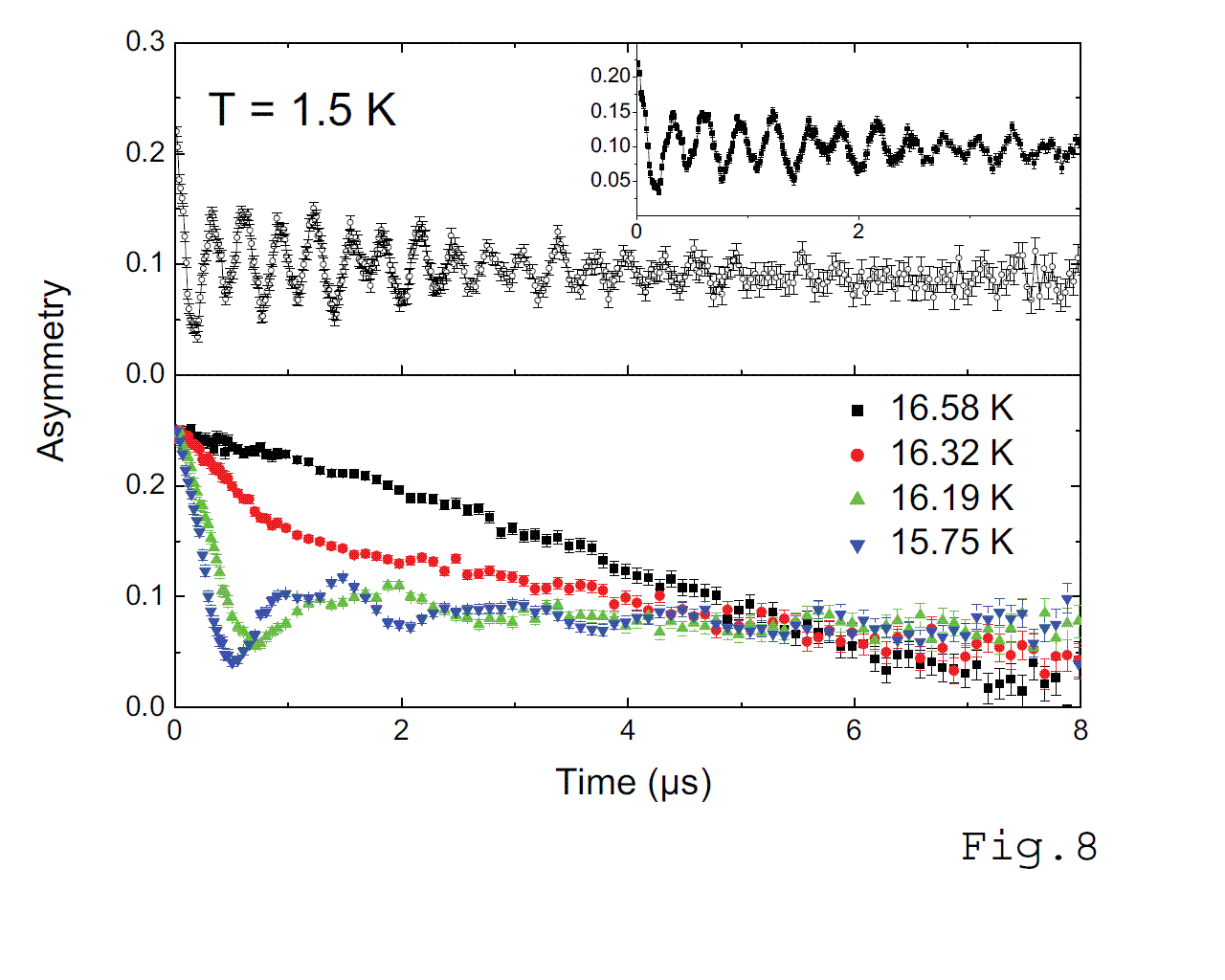';file-properties
"XNPEU";}}

\FRAME{ftbpF}{7.3734in}{5.22in}{0pt}{}{}{figure9.png}{\special{language
"Scientific Word";type "GRAPHIC";maintain-aspect-ratio TRUE;display
"ICON";valid_file "F";width 7.3734in;height 5.22in;depth 0pt;original-width
23.3266in;original-height 16.4868in;cropleft "0";croptop "1";cropright
"1";cropbottom "0";filename '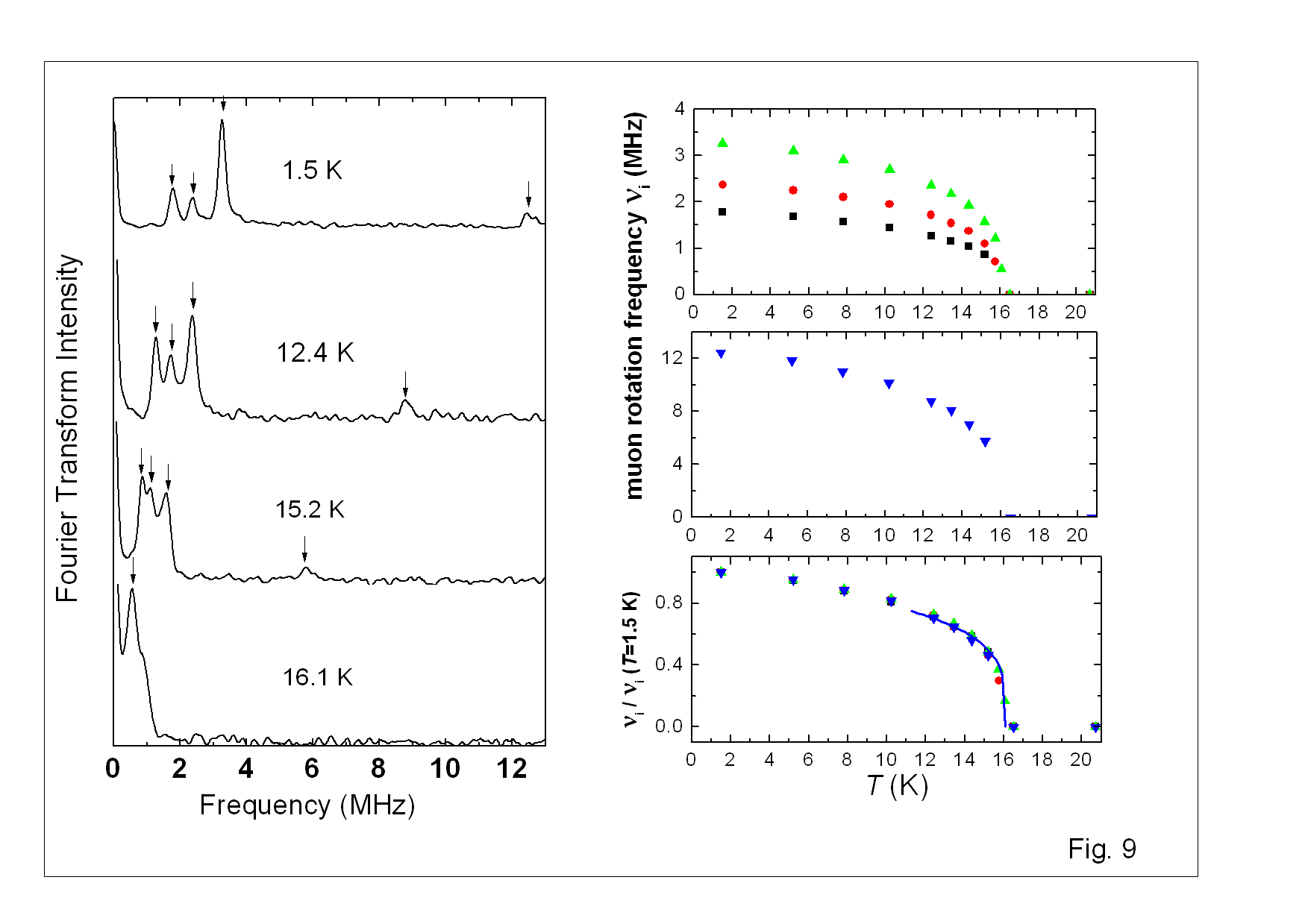';file-properties "XNPEU";}}

\FRAME{ftbpF}{4.0395in}{5.22in}{0pt}{}{}{figure10.png}{\special{language
"Scientific Word";type "GRAPHIC";maintain-aspect-ratio TRUE;display
"ICON";valid_file "F";width 4.0395in;height 5.22in;depth 0pt;original-width
16.9668in;original-height 21.9533in;cropleft "0";croptop "1";cropright
"1";cropbottom "0";filename '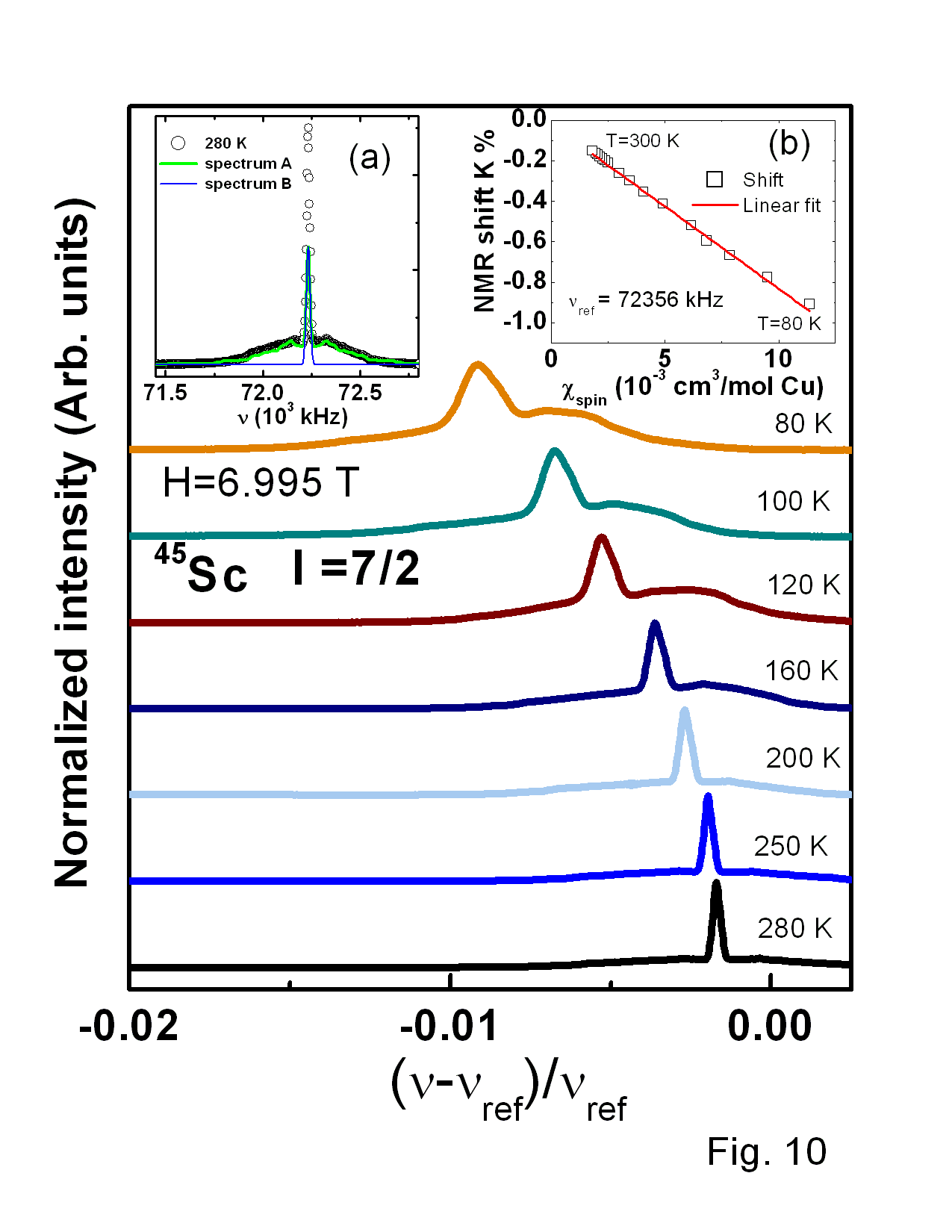';file-properties "XNPEU";}}

\FRAME{ftbpF}{7.2566in}{5.2209in}{0pt}{}{}{figure11.png}{\special{language
"Scientific Word";type "GRAPHIC";maintain-aspect-ratio TRUE;display
"ICON";valid_file "F";width 7.2566in;height 5.2209in;depth
0pt;original-width 22.2265in;original-height 15.967in;cropleft "0";croptop
"1";cropright "1";cropbottom "0";filename '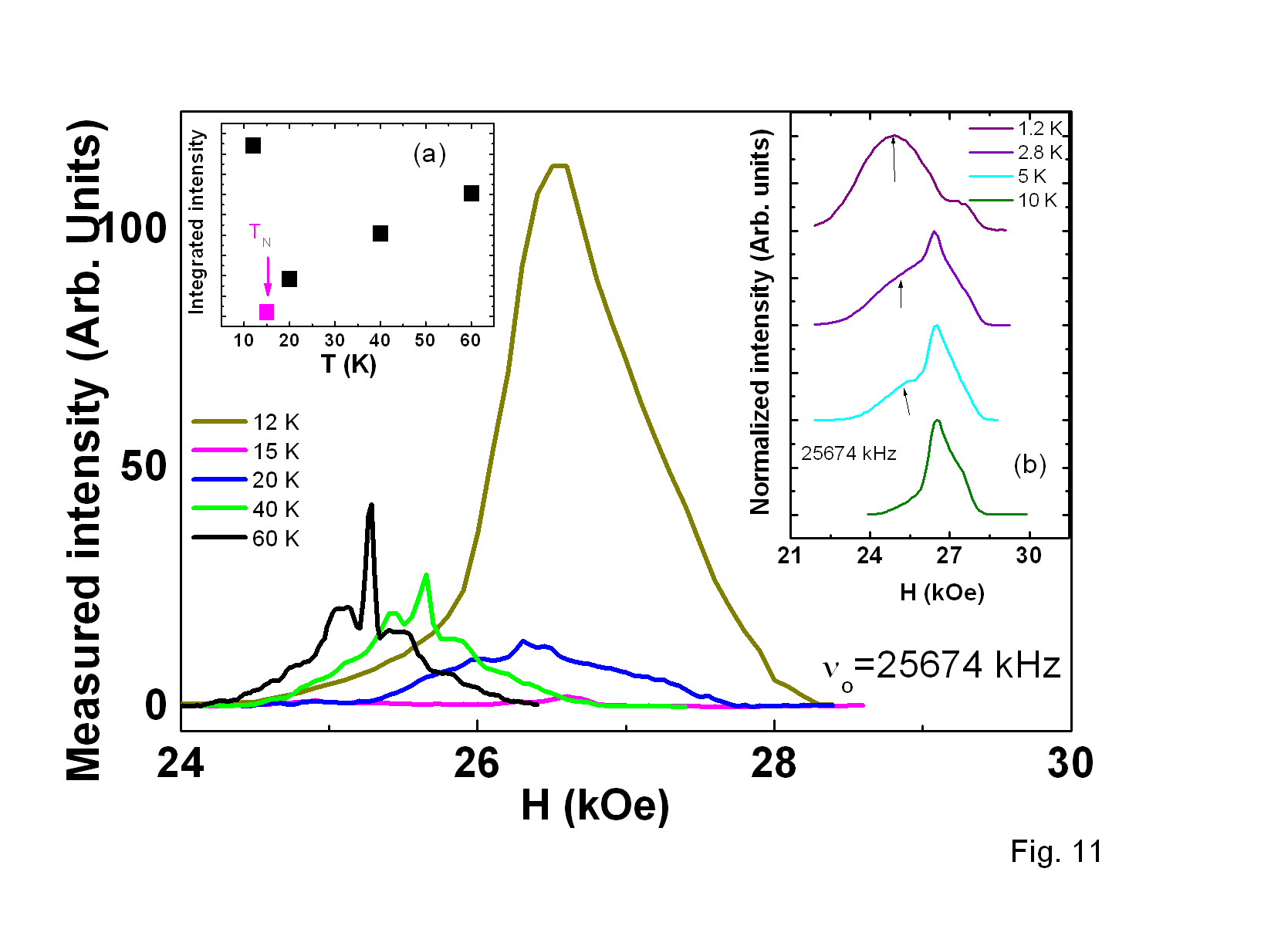';file-properties
"XNPEU";}}

\FRAME{ftbpF}{7.2566in}{5.2209in}{0pt}{}{}{figure12.png}{\special{language
"Scientific Word";type "GRAPHIC";maintain-aspect-ratio TRUE;display
"ICON";valid_file "F";width 7.2566in;height 5.2209in;depth
0pt;original-width 22.2265in;original-height 15.967in;cropleft "0";croptop
"1";cropright "1";cropbottom "0";filename '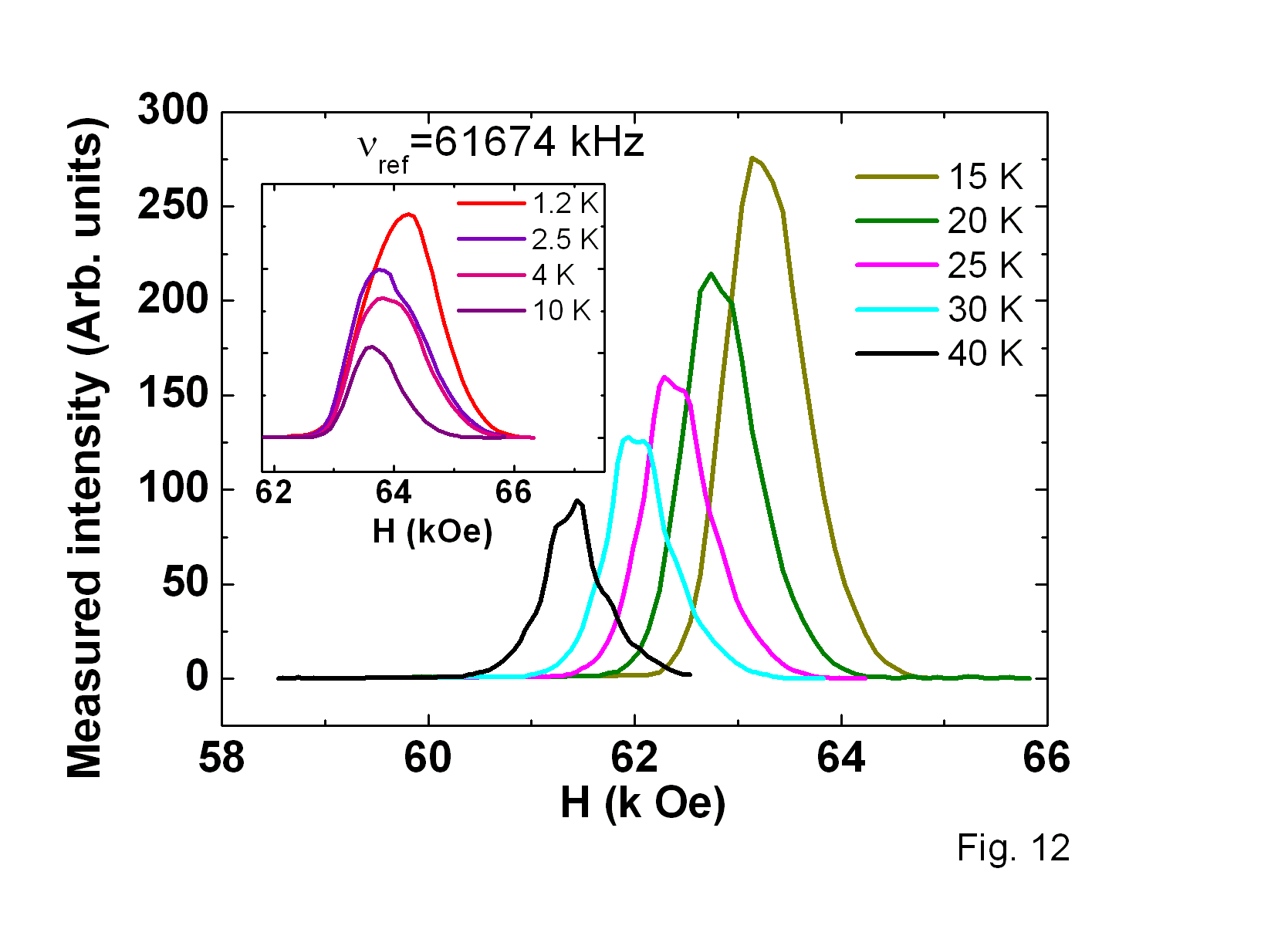';file-properties
"XNPEU";}}

\FRAME{ftbpF}{6.8087in}{5.22in}{0pt}{}{}{figure13.png}{\special{language
"Scientific Word";type "GRAPHIC";maintain-aspect-ratio TRUE;display
"ICON";valid_file "F";width 6.8087in;height 5.22in;depth 0pt;original-width
21.3133in;original-height 16.3207in;cropleft "0";croptop "1";cropright
"1";cropbottom "0";filename '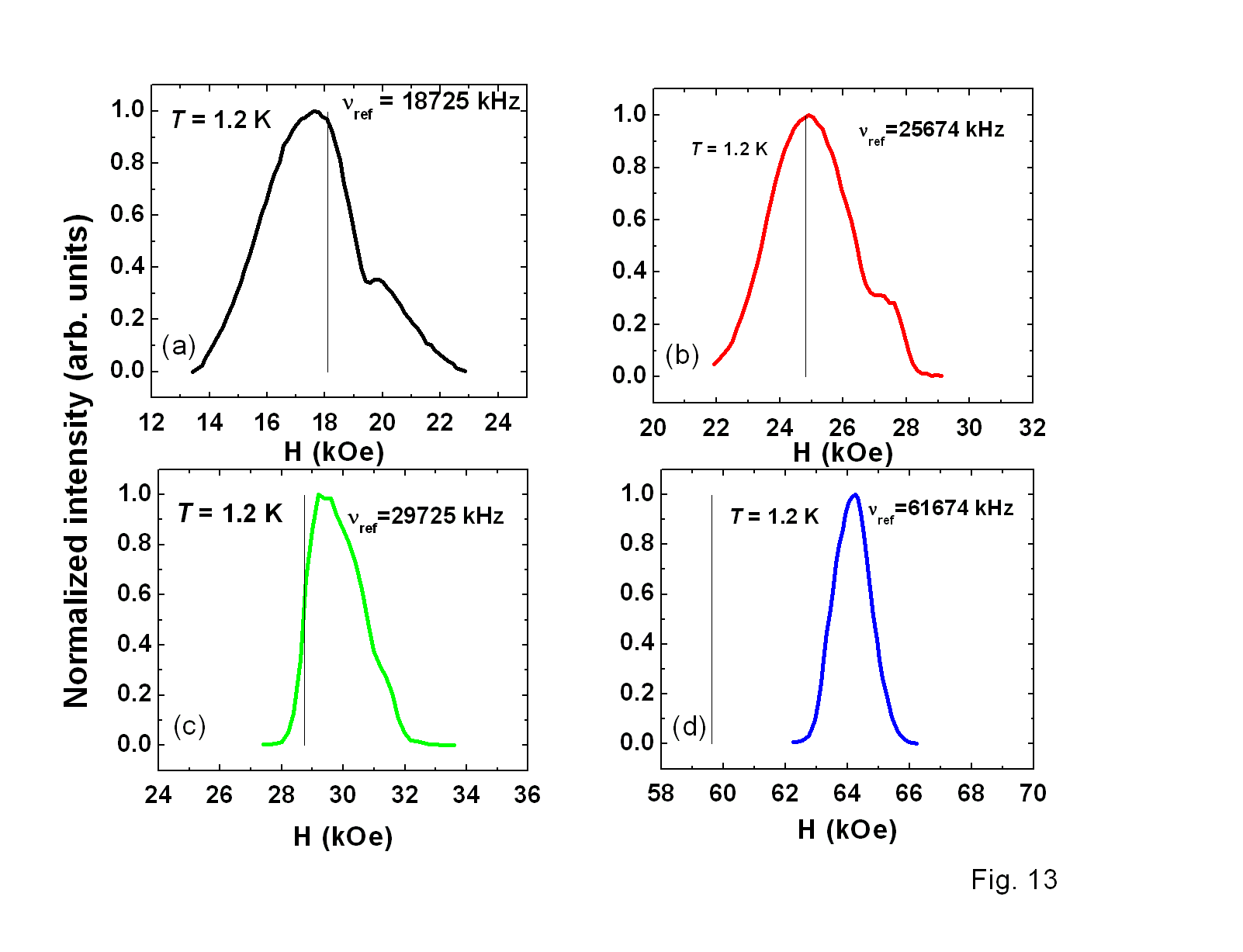';file-properties "XNPEU";}}

\FRAME{ftbpF}{5.8859in}{5.2226in}{0pt}{}{}{figure14.png}{\special{language
"Scientific Word";type "GRAPHIC";maintain-aspect-ratio TRUE;display
"ICON";valid_file "F";width 5.8859in;height 5.2226in;depth
0pt;original-width 7.6994in;original-height 6.826in;cropleft "0";croptop
"1";cropright "1";cropbottom "0";filename '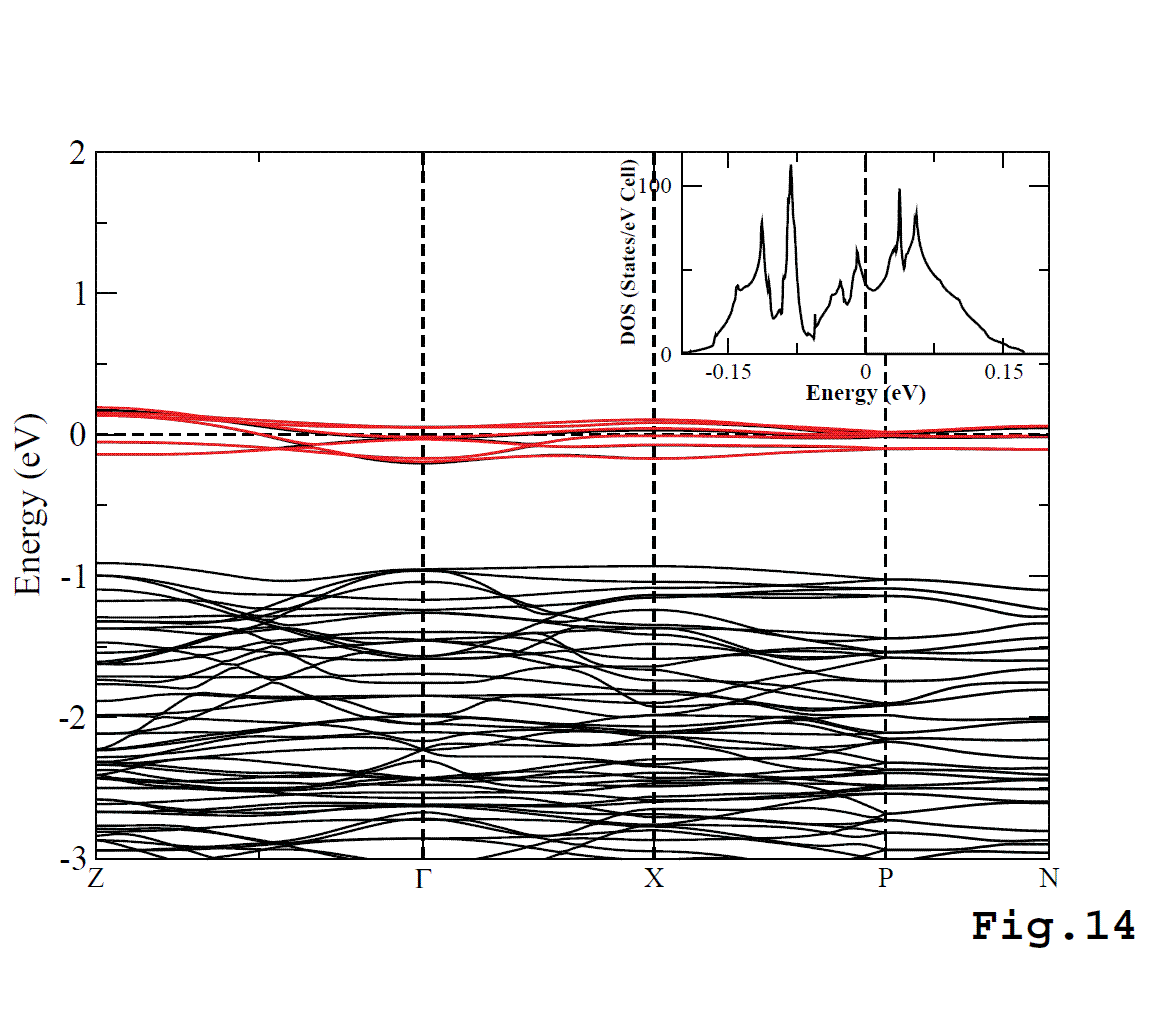';file-properties
"XNPEU";}}

\FRAME{ftbpF}{7.6752in}{5.892in}{0pt}{}{}{figure15.png}{\special{language
"Scientific Word";type "GRAPHIC";maintain-aspect-ratio TRUE;display
"ICON";valid_file "F";width 7.6752in;height 5.892in;depth 0pt;original-width
7.6069in;original-height 5.8332in;cropleft "0";croptop "1";cropright
"1";cropbottom "0";filename '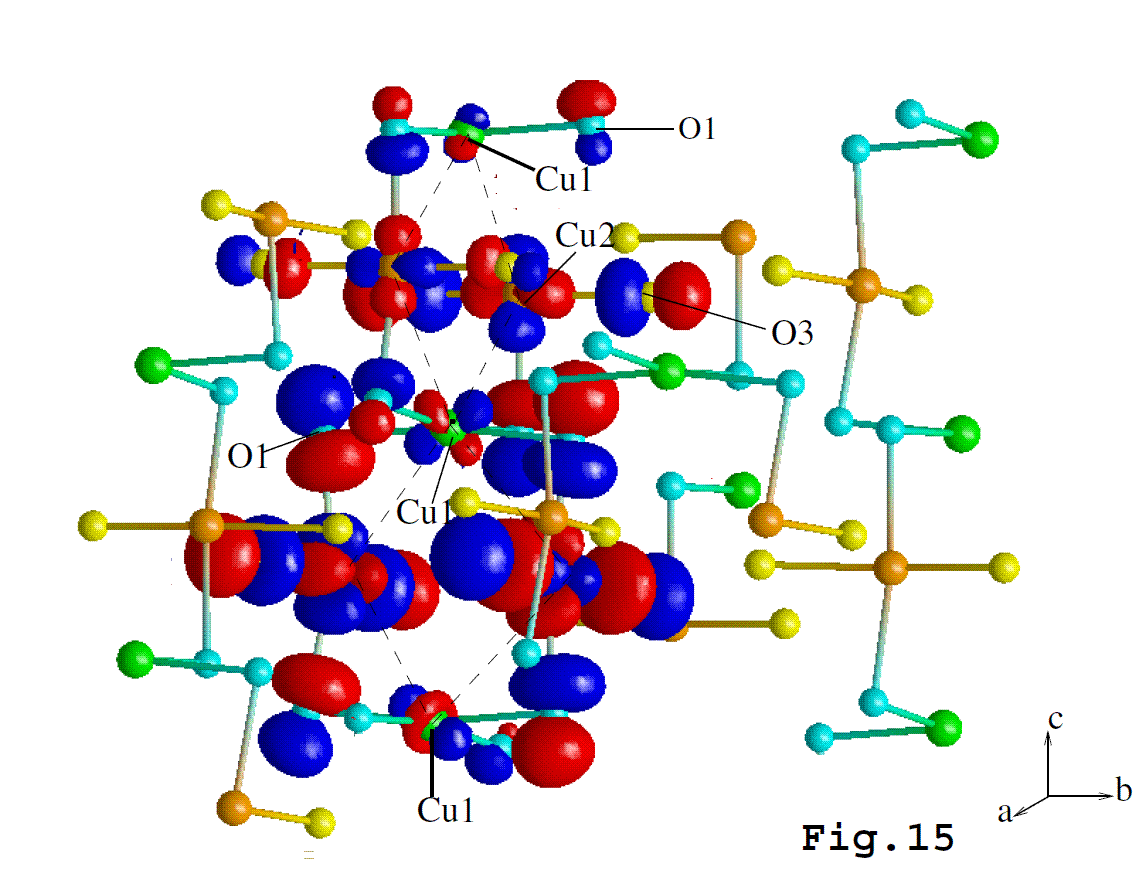';file-properties "XNPEU";}}

\end{document}